\documentclass[]{article}
\usepackage[left=2.2cm,right=2.2cm,top=2cm,bottom=2cm]{geometry}

\usepackage[utf8]{inputenc}
\usepackage{amsmath, amsthm, amssymb}
\usepackage{graphicx}
\usepackage{enumitem}
\usepackage[justification=centering,labelfont=bf]{caption}
\usepackage{url}
\usepackage{multicol}
\usepackage{setspace}
\usepackage{placeins}
\usepackage{pifont}

\usepackage{indentfirst}
\usepackage[noline, noend, nofillcomment, linesnumbered]{algorithm2e}
\SetArgSty{textnormal}
\SetCommentSty{texttt}
\SetNoFillComment
\SetNlSty{textnormal}{}{}

\SetAlgoCaptionLayout{Centerline}
\SetNlSkip{0.8em}
\SetNlSty{texttt}{}{}
\let\oldnl\nl% Store \nl in \oldnl
\newcommand{\nonl}{\renewcommand{\nl}{\let\nl\oldnl}}% Remove line number for one line

\setlist{nosep}

\newtheorem{definition}{Definition}

\newcommand*{\Xbar}[1]{\overline{#1}}

\newcommand{\Xund}{\rule{.5em}{.5pt}}

\newcommand{\XF}{\mathcal{F}}

\newcommand{\YN}{\mathbb{N}}
\newcommand{\YO}{\mathbb{O}}

\begin{document}

\title{A closer look at TDFA}
%\title{TDFA -- Fast Submatch Extraction in Regular Expressions}
%\title{Fast submatch extraction with lookahead-TDFA}
\author{
    Angelo Borsotti \\
    \texttt{\small{angelo.borsotti@mail.polimi.it}}
\and
    Ulya Trafimovich \\
    \texttt{\small{skvadrik@gmail.com}}
}
\date{2022}

\maketitle

\begin{abstract}
We present an algorithm for regular expression parsing and submatch extraction
based on tagged deterministic finite automata.
The algorithm works with different disambiguation policies.
We give detailed pseudocode for the algorithm, covering important practical optimizations.
All transformations from a regular expression to an optimized automaton are explained on a step-by-step example.
We consider both ahead-of-time and just-in-time determinization
and describe variants of the algorithm suited to each setting.
We provide benchmarks showing that the algorithm is very fast in practice.
Our research is based on two independent implementations:
an open-source lexer generator RE2C
and an experimental Java library.
\end{abstract}

\section*{Introduction}
This paper describes tagged deterministic finite automata (TDFA).
To the best of our knowledge, it is the first \emph{practical} submatch extraction and parsing algorithm based on deterministic finite automata
that is capable of both POSIX and leftmost greedy disambiguation.
%It is primarily targeted at practical readers who want to implement regular expression parsing or submatch extraction
%in a lexer generator, a runtime library or some other regular expression engine.
Most of the theory behind TDFA is not new, but the previous papers are incomplete and lack important details.
This paper consolidates our previous research
and provides a comprehensive description of the algorithm.
We hope that it will make TDFA easier to implement in practice.
\medskip

Here is a brief history of TDFA development.
In 2000 Laurikari published the original paper \cite{Lau00}.
In 2007 Kuklewicz implemented TDFA in a Haskell library with POSIX longest-match disambiguation;
he gave only an informal description \cite{Kuk07}.
In 2016 Trafimovich presented TDFA with lookahead \cite{Tro17},
implemented them in the open-source lexer generator RE2C \cite{RE2C}
and formalized Kuklewicz disambiguation algorithm.
In 2017 Borsotti implemented TDFA in an experimental Java library \cite{RE2CJava}.
In 2019 Borsotti and Trafimovich adapted Okui-Suzuki disambiguation algorithm to TDFA
and showed it to be faster than Kuklewicz algorithm \cite{BorTro19}.
In 2020 Trafimovich published an article about TDFA implementation in RE2C \cite{Tro20}.
Finally, the present paper incorporates our past research
and adds novel findings on multi-pass TDFA that are better suited to just-in-time determinization.
\medskip

Before diving into details, we recall the key concepts discussed in the paper.
\medskip

\emph{Regular expressions} (REs) are a notation for describing sets of strings known as regular languages, or Type-3 languages in the Chomsky hierarchy.
They were first defined by Kleene \cite{Kle51} as sets of strings constructed from the alphabet symbols and the empty word via the application of three basic operations: concatenation, alternative and iteration.
Later REs were formalized via the notion of Kleene algebra \cite{Koz94}.
In practice REs have many extensions that vary in complexity and expressive power.
\medskip

\emph{Submatch extraction} is an extension of REs that allows one to
identify input positions matching specific positions in a RE.
Recall the difference between \emph{recognition} and \emph{parsing}:
to recognize a string means to determine its membership in a language,
but to parse a string means to also find its derivation in a language grammar.
Submatch extraction is in between:
on one extreme it approaches recognition (if there are no submatch markers in a RE),
but on the other extreme it is identical to parsing (if every position in a RE is marked).
In general it requires constructing a partial derivation,
which can be implemented more efficiently than full parsing.
\medskip

\emph{Finite state automata} are a formalism equivalent to REs
in the sense that every RE can be converted to a deterministic finite automaton (DFA)
or a nondeterministic finite automaton (NFA), and vice versa.
There are many different kinds of NFA, but there is a unique minimal DFA for a given RE.
Both NFA and DFA solve the recognition problem for REs in linear time in the length of input.
In practice DFA are faster because they follow a single path,
while NFA have to track multiple paths simultaneously.
NFA can be converted to DFA using a \emph{determinization} procedure,
but the resulting DFA may be exponentially larger than the NFA.
\medskip

\emph{Tags} are submatch markers: they mark positions in a RE that should be mapped to offsets in the input string.
When a RE is converted to an NFA, tags are placed on the NFA transitions.
This effectively turns NFA into a nondeterministic finite-state \emph{transducer}
that rewrites symbolic strings into tagged strings (where tags are placed in-between symbols, marking submatch boundaries).
Conversion from a RE to a tagged NFA is natural if NFA mirrors the structure of RE, as in the case of Thompson's construction.
\medskip

Determinization of a tagged NFA is problematic, because in a DFA multiple NFA paths are collapsed into one,
causing conflicts when the same tag has different values on different NFA paths.
To keep track of all possible tag values, a DFA is augmented with \emph{registers}
and \emph{operations} on transitions that update register values.
The number of registers and operations depends only on the RE structure and tag density,
but not on the input string,
therefore it adds only a constant overhead to the DFA execution.
We describe techniques that reduce redundant operations and minimize the overhead in practice.
\medskip

\emph{Ambiguity} is yet another problem for submatch extraction;
it means the existence of multiple different parse trees for the same input.
Ambiguity should not be confused with non-determinism,
which means the existence of multiple possibilities during matching that get canceled as more input is consumed;
ambiguity is a property of a RE.
One way to resolve it is a \emph{disambiguation policy},
the most notable examples being the leftmost-greedy and the longest-match (POSIX) policies.
TDFA can work with both policies, and there is no runtime overhead on disambiguation --- it is built into TDFA structure.
%POSIX disambiguation is covered in a separate paper \cite{BorTro19}.
Some RE engines provide other ways to resolve ambiguity, such as user-defined precedence rules,
but these are ad-hoc, error-prone and often difficult to reason about.
\medskip

RE engines based on DFA can be divided in two groups:
those using \emph{ahead-of-time} (AOT) determinization (e.g. lexer generators)
and those using \emph{just-in-time} (JIT) determinization (e.g. runtime libraries).
The former can spend considerable amount of time on preprocessing,
but the latter face a tradeoff between the time spent on preprocessing and the time spent on matching.
Therefore it makes sense to use different variants of the algorithm in each case.
We describe \emph{single-pass} TDFA that are a natural fit for ahead-of-time determinization,
and \emph{multi-pass} TDFA that are better suited to just-in-time determinization.
\medskip

In practice performance of a matching algorithm depends on the representation of results.
The most generic representation is a parse tree; it precisely reflects a derivation.
A more lightweight representation is a list of offsets or a single offset per submatch position in a RE
(the latter is used in the POSIX \texttt{regexec} function).
Another representation, more suitable for transducers, is a \emph{tagged string} ---
a sequence of input symbols interspersed with tags.
If a RE contains tags for every subexpression,
then it is possible to reconstruct a parse tree from offset lists or a tagged string (a procedure is given in \cite{BorTro19}, section 6).
TDFA can be used with all the above representations,
but it is more natural to use offsets with single-pass TDFA
and tagged strings with multi-pass TDFA.
\medskip

The rest of the paper is structured as follows.
Section \ref{section_tnfa} defines REs and their conversion to nondeterministic automata.
Section \ref{section_tdfa} defines TDFA and determinization.
Section \ref{section_impl} describes optimizations and practical implementation details.
Section \ref{section_multipass} describes multi-pass TDFA and their application to just-in-time determinization.
Section \ref{section_evaluation} provides benchmarks and comparison with other algorithms.
Section \ref{section_conclusions} contains conclusions,
and section \ref{section_future_work} contains ideas for future work.

\section*{Conventions}

In this paper we use \emph{pseudocode} rather than formal mathematical notation to describe algorithms.
We focus on the practical side, because we want to encourage TDFA implementation in real-world programs.
The most theoretically challenging part of the algorithm (POSIX disambiguation)
is formalized in our previous paper \cite{BorTro19},
and the core of the algorithm is based on the well-known idea of determinization via powerset construction
that does not need a formal introduction.
\medskip

In the pseudocode, we try to balance between formality and clarity.
We omit the definitions of basic operations on data structures, such as ``append to a list'' or ``push on stack''.
We sometimes use set notation with predicates, and sometimes prefer explicit loops that iterate over the elements of a set.
To reduce verbosity, we assume that function arguments are passed by reference and modifications to them are visible in the calling function
(although some functions have explicit return values).
\medskip

All algorithms presented here are implemented in the open-source lexer generator RE2C and are known to work in practice.

\pagebreak

\section{TNFA}\label{section_tnfa}

%In this section we formalize TDFA and prove that the algorithm is correct.
%We do that in the following steps.
%First, we formalize RE.
%Then we show how to convert RE to NFA that closely mirrors the structure of RE, and therefore preserves submatch information and ambiguity in it.
%Then we show how to simulate NFA and obtain submatch results.
%Since NFA has the same structure and submatch information as the original RE,
%and since the simulation procedure is a trivial modification of the canonical simulation,
%it is easy to see that it yields correct submatch values.
%The only tricky part is disambiguation, but in this paper we take it for granted ---
%POSIX disambiguation is described in great detail in another paper \cite{BoRTro19},
%and leftmost greedy disambiguation is trivial since by definition it can be implemented with a simple leftmost depth-first search over the NFA.
%Then we show how to convert NFA to DFA and how to execute DFA.
%Finally, we show that DFA execution yields the same results as NFA simulation, which proves that the algorithm is correct.

In this section we define regular expressions, their conversion to nondeterministic automata and matching.

\begin{definition}
Regular expressions (REs) over finite alphabet $\Sigma$ are:
\begin{enumerate}
    \item
      Empty RE $\epsilon$,
      unit RE $a \in \Sigma$
      and tag $t \in \YN$.
    \item Alternative $e_1 | e_2$,
      concatenation $e_1 e_2$ and
      repetition $e_1^{n, m}$ $(0 \!\leq\! n \!\leq\! m \!\leq\! \infty)$
      where $e_1$ and $e_2$ are REs over $\Sigma$.
\end{enumerate}
\end{definition}

Tags mark submatch positions in REs.
Unlike capturing parentheses, tags are not necessarily paired.
Capturing groups can be represented with tags,
but the correspondence may be more complex than a pair of tags per group,
e.g. POSIX capturing groups require additional hierarchical tags \cite{BorTro19}.
\medskip

Generalized repetition $e^{n, m}$ can be bounded ($m < \infty$) or unbounded ($m = \infty$).
Unbounded repetition $e^{0,\infty}$ is the canonical Kleene iteration, shortened as $e^*$.
Bounded repetition is usually desugared via concatenation,
but we avoid desugaring as it may duplicate tags and change submatch semantics in a RE.

\begin{definition} \label{def_tnfa}
Tagged Nondeterministic Finite Automaton (TNFA)
is a structure $(\Sigma, T, Q, q_0, q_f, \Delta)$, where:
\begin{itemize}
    \item[] $\Sigma$ is a finite set of symbols (alphabet)
    \item[] $T$ is a finite set of tags
    \item[] $Q$ is a finite set of states with initial state $q_0$ and final state $q_f$
    \item[] $\Delta$ is a transition relation that contains transitions of two kinds:
    \begin{itemize}
        \item[] transitions on alphabet symbols $(q, a, p)$ where $q, p \in Q$ and $a \in \Sigma$
        \item[] optionally tagged $\epsilon$-transitions with priority $(q, i, t, p)$ where $q, p \in Q$, $i \in \YN$ and $t \in T \cup \Xbar{T} \cup \{\epsilon\}$
    \end{itemize}
\end{itemize}
\end{definition}

TNFA is in essence a non-deterministic finite state transducer with input alphabet $\Sigma$ and output alphabet $\Sigma \cup T \cup \Xbar{T}$.
$\Xbar{T} = \{-t \mid t \in T\}$ is the set of all negative tags which represent the absence of match: they appear whenever there is a way to bypass a tagged subexpression in a RE,
such as alternative or repetition with zero lower bound.
Negative tags serve a few purposes:
they prevent propagation of stale submatch values from one iteration to another,
they spares the need to initialize tags,
and they are needed for POSIX disambiguation \cite{BorTro19}.
Priorities are used for transition ordering during $\epsilon$-closure construction.
\medskip

Algorithm \ref{alg_tnfa} on page \pageref{alg_tnfa} shows TNFA construction:
it performs top-down structural recursion on a RE, passing the final state on recursive descent into subexpressions
and using it to connect subautomata.
This is similar to Thompson's construction, except that non-essential $\epsilon$-transitions are removed and tagged transitions are added.
The resulting automaton mirrors the structure of RE and preserves submatch information and ambiguity in it.

\begin{algorithm}[] \DontPrintSemicolon \SetKwProg{Fn}{}{}{} \SetAlgoInsideSkip{medskip}
\begin{multicols}{2}
\setstretch{0.9}
\small
\Indm

\nonl\Fn {$\underline{simulation \big( (\Sigma, T, Q, q_0, q_f, \Delta), \; a_1 \hdots a_n \big)} \smallskip$} {
    $m_0:$ vector of offsets of size $|T|$ \;
    $C = \{ (q_0, m_0) \}$ \;
    \BlankLine
    \For {$k = \overline{1, n}$} {
        $C = epsilon \Xund closure(C, \Delta, q_f, k)$ \;
        $C = step \Xund on \Xund symbol(C, \Delta, a_k)$ \;
        \lIf {$C = \emptyset$} {
            \Return $\varnothing$
        }
    }
    \BlankLine
    $C = epsilon \Xund closure(C, \Delta, q_f, n)$ \;
    \BlankLine
    \lIf {$\exists (q, m)$ in $C \mid q = q_f$} {
        \Return $m$
    } \lElse {
        \Return $\varnothing$
    }
}
\vspace{2em}

\nonl\Fn {$\underline{step \Xund on \Xund symbol \big( C, \Delta, a \big)} \smallskip$} {
    \Return $\{ (p, m) \mid (q, m)$ in $C$ and $(q, a, p) \in \Delta \}$ \;
}

\vfill\null
\columnbreak

\nonl\Fn {$\underline{epsilon \Xund closure \big( C, \Delta, q_f, k \big)} \smallskip$} {
    $C':$ empty sequence of configurations \;
    \BlankLine
    \For {$(q, m)$ in $C$ in reverse order} {
        push $(q, m)$ on stack \;
    }
    \BlankLine
    \While {stack is not empty} {
        pop $(q, m)$ from stack \;
        append $(q, m)$ to $C'$ \;
        \BlankLine
        \For {each $(q, i, t, p) \in \Delta$ ordered by priority $i$} {
            \lIf {$t > 0$} {
                $m[t] = k$
            } \lElse {
                $m[-t] = \mathbf{n}$
            }
            %\BlankLine
            \If {configuration with state $p$ is not in $C'$} {
                push $(p, m)$ on stack \;
            }
        }
    }
    \BlankLine
    \Return $\{ (q, m)$ in $C' \mid q = q_f$ or \\
        \hphantom{\Return $\{ (q, m)$ in $C' \mid$} $\exists (q, a, \Xund) \in \Delta$ where $a \in \Sigma \}$  \;
}

\vfill\null

\end{multicols}
%\vspace{1em}
\caption{TNFA simulation.}\label{alg_simulation}
\end{algorithm}

Algorithm \ref{alg_simulation} defines TNFA simulation on a string.
It starts with a single configuration $(q_0, m_0)$ consisting of the initial state $q_0$ and an empty vector of tag values,
and loops over the input symbols until all of them are matched or the configuration set becomes empty, indicating match failure.
At each step the algorithm constructs $\epsilon$-closure of the current configuration set, updating tag values along the way, and steps on transitions labeled with the current input symbol.
Finally, if all symbols have been matched and there is a configuration with the final state $q_f$, the algorithm terminates successfully and returns the final vector of tag values.
Otherwise it returns a failure.
The algorithm uses leftmost greedy disambiguation; POSIX disambiguation is more complex and requires a different $\epsilon$-closure algorithm \cite{BorTro19}.
Figure \ref{fig:tdfa} in section \ref{section_tdfa} shows an example of TNFA simulation.

%\FloatBarrier

\begin{algorithm}[] \DontPrintSemicolon \SetKwProg{Fn}{}{}{}
\begin{multicols}{2}

    \newcommand \retonfa {tn\!f\!a}%{\XN}
    \newcommand \ntag {ntags}
    \Indm
    \small
    \setstretch{1.1}

    \nonl\Fn {$\underline{\retonfa(e, q_f)} \smallskip$} {
    %\Indm

    \If {$e = \epsilon$} {
        \Return $(\Sigma, \emptyset, \{q_f\}, q_f, q_f, \emptyset)$
    }
    \BlankLine
    \BlankLine

    \ElseIf {$e = a \in \Sigma$} {
        \Return $(\Sigma, \emptyset, \{q_0,q_f\}, q_0, q_f, \{(q_0, a, q_f)\})$
    }
    \BlankLine
    \BlankLine

    \ElseIf {$e = t \in \YN$} {
        \Return $(\Sigma, \{t\}, \{q_0, q_f\}, q_0, q_f, \{(q_0, 1, t, q_f)\})$
    }
    \BlankLine
    \BlankLine

    \ElseIf {$e = e_1 \cdot e_2$} {
        $(\Sigma, T_2, Q_2, q_2, q_f, \Delta_2) = \retonfa (e_2, q_f)$ \;
        $(\Sigma, T_1, Q_1, q_1, q_2, \Delta_1) = \retonfa (e_1, q_2)$ \;
        \Return $(\Sigma, T_1 \cup T_2, Q_1 \cup Q_2, q_1, q_f, \Delta_1 \cup \Delta_2)$
    }
    \BlankLine
    \BlankLine

    \ElseIf {$e = e_1 \mid e_2$} {
        $(\Sigma, T_2, Q_2, q_2, q_f, \Delta_2) = \retonfa (e_2, q_f)$ \;
        $(\Sigma, T_2, Q'_2, q'_2, q_f, \Delta'_2) = \ntag (T_2, q_f)$ \;
        $(\Sigma, T_1, Q_1, q_1, q'_2, \Delta_1) = \retonfa (e_1, q'_2)$ \;
        $(\Sigma, T_1, Q'_1, q'_1, q_2, \Delta'_1) = \ntag (T_1, q_2)$ \;
        $Q = Q_1 \cup Q'_1 \cup Q_2 \cup Q'_2 \cup \{q_0\}$ \;
        $\Delta = \Delta_1 \cup \Delta'_1 \cup \Delta_2 \cup \Delta'_2 \cup \{ (q_0,1,\epsilon,q_1), (q_0,2,\epsilon,q'_1) \}$ \;
        \Return $(\Sigma, T_1 \cup T_2, Q, q_0, q_f, \Delta)$
    }
    \BlankLine
    \BlankLine

    \ElseIf {$e = e_1^{n, m} \mid_{1 < n \leq m \leq \infty}$} {
        $(\Sigma, T_1, Q_1, q_2, q_f, \Delta_1) = \retonfa(e_1^{n\!-\!1, m\!-\!1}, q_f)$ \;
        $(\Sigma, T_2, Q_2, q_1, q_2, \Delta_2) = \retonfa(e_1, q_2)$ \;
        \Return $(\Sigma, T_1 \cup T_2, Q_1 \cup Q_2, q_1, q_f, \Delta_1 \cup \Delta_2)$
    }
    \BlankLine
    \BlankLine

    \ElseIf {$e = e_1^{1, m} \mid_{1 < m < \infty}$} {
        \lIf {$m = 1$} {
            \Return $\retonfa (e_1, q_f)$
        }
        $(\Sigma, T_1, Q_1, q_1, q_f, \Delta_1) = \retonfa (e_1^{1, m\!-\!1}, q_f)$ \;
        $(\Sigma, T_2, Q_2, q_0, q_2, \Delta_2) = \retonfa (e_1, q_1)$ \;
        $\Delta = \Delta_1 \cup \Delta_2 \cup \{ (q_1, 1, \epsilon, q_f), (q_1, 2, \epsilon, q_2) \}$ \;
        \Return $(\Sigma, T_1 \cup T_2, Q_1 \cup Q_2, q_0, q_f, \Delta)$
    }
    \BlankLine
    \BlankLine

    \ElseIf {$e = e_1^{0, m}$} {
        $(\Sigma, T_1, Q_1, q_1, q_f, \Delta_1) = \retonfa (e_1^{1, m}, q_f)$ \;
        $(\Sigma, T_1, Q'_1, q'_1, q_f, \Delta'_1) = \ntag (T_1, q_f)$ \;
        $Q = Q_1 \cup Q'_1 \cup \{q_0\}$ \;
        $\Delta = \Delta_1 \cup \Delta'_1 \cup \{ (q_0, 1, \epsilon, q_1), (q_0, 2, \epsilon, q'_1) \}$ \;
        \Return $(\Sigma, T_1, Q, q_0, q_f, \Delta)$
    }
    \BlankLine
    \BlankLine

    \ElseIf {$e = e_1^{1, \infty}$} {
        $(\Sigma, T_1, Q_1, q_0, q_1, \Delta_1) = \retonfa (e_1, q_1)$ \;
        $Q = Q_1 \cup \{q_f\}$ \;
        $\Delta = \Delta_1 \cup \{ (q_1, 1, \epsilon, q_0), (q_1, 2, \epsilon, q_f) \}$ \;
        \Return $(\Sigma, T_1, Q, q_0, q_f, \Delta)$
    }
    }
    \vspace{2em}

    \nonl\Fn {$\underline{\ntag(T, q_f)} \smallskip$} {
        \Indp
        $\{ t_i \}_{i=1}^n = T$ \;
        $Q = \{q_i\}_{i=0}^n$ where $q_n = q_f$ \;
        $\Delta = \{ (q_{i-1}, 1, -t_i, q_i) \}_{i=1}^n$ \;
        \Return $(\Sigma, T, Q, q_0, q_f, \Delta)$ \;
    }

    \vfill\null

\columnbreak

    \nonl \includegraphics[width=\linewidth]{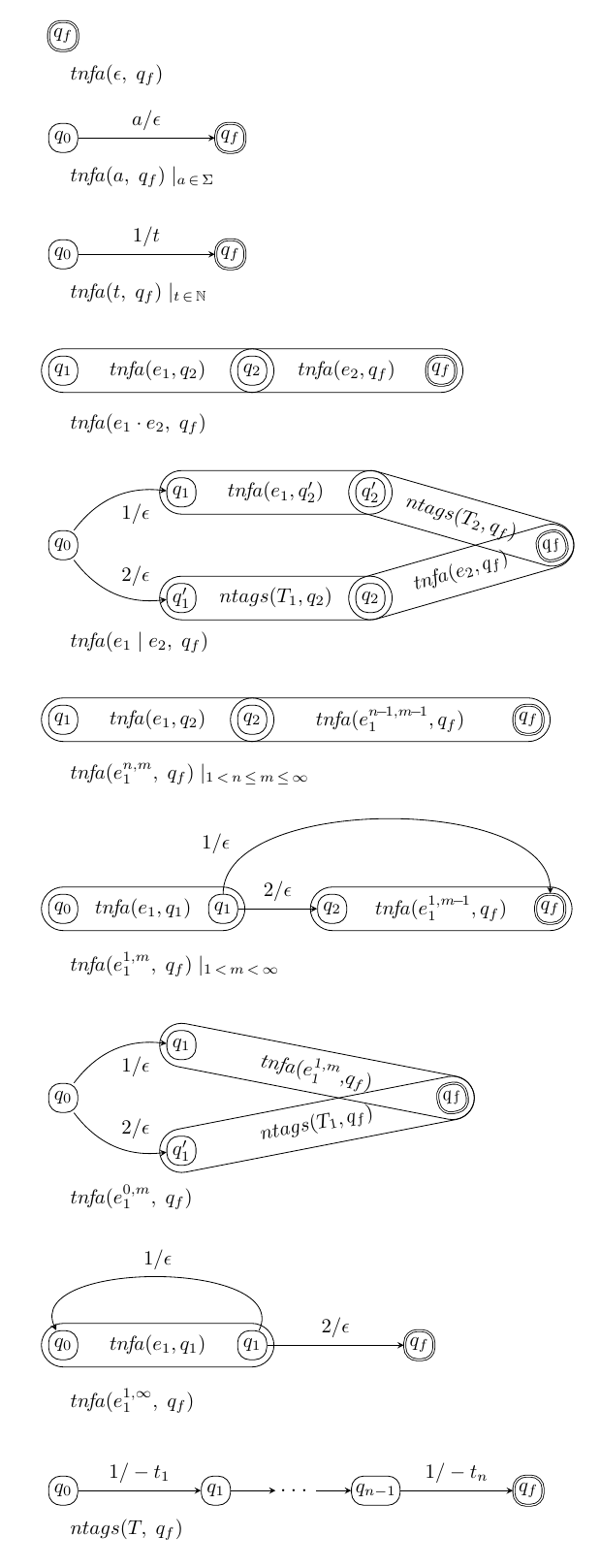}

\end{multicols}
\vspace{1em}
\caption{TNFA construction.}\label{alg_tnfa}
\end{algorithm}

\pagebreak

\section{TDFA}\label{section_tdfa}

In this section we define TDFA and show how to convert TNFA to TDFA.

\begin{definition} \label{def_tnfa}
Tagged Deterministic Finite Automaton (TDFA)
is a structure $(\Sigma, T, S, S_f, s_0, R, R_f, \delta, \varphi)$, where:
\begin{itemize}
    \item[] $\Sigma$ is a finite set of symbols (alphabet)
    \item[] $T$ is a finite set of tags
    \item[] $S$ is a finite set of states with initial state $s_0$ and a subset of final states $S_f \subseteq S$
    \item[] $R$ is a finite set of registers with a subset of final registers $R_f$ (one per tag)
    \item[] $\delta : S \times \Sigma \rightarrow S \times \YO^*$ is a transition function
    \item[] $\varphi : S_f \rightarrow \YO^*$ is a final function
    \medskip
    \item[] where $\YO$ is a set of register operations of the following types:
    \begin{itemize}
        \item[] set register $i$ to nil or to the current position: $i \leftarrow v$, where $v \in \{\mathbf{n}, \mathbf{p}\}$
        \item[] copy register $j$ to register $i$: $i \leftarrow j$
        \item[] copy register $j$ to register $i$ and append history: $i \leftarrow j \cdot h$, where $h$ is a string over $\{\mathbf{n}, \mathbf{p}\}$
    \end{itemize}
\end{itemize}
\end{definition}

Compared to an ordinary DFA, TDFA is extended with a set of tags $T$,
a set of registers $R$ with one final register per tag,
and register operations that are attributed to transitions and final states (the $\delta$ and $\varphi$ functions).
$\YO^*$ denotes the set of all sequences of operations over $\YO$.
Operations can be of three types: set, copy, append.
Set operations are used for \emph{single-valued} tags (those represented with a single offset),
append operations are used for \emph{multi-valued} tags (those represented with an offset list), and
copy operations are used for all tags.
The decision which tags are single-valued and which ones are multi-valued is arbitrary and individual for each tag
(it may be, but does not have to be based on whether the tag is under repetition).
Register values are denoted by special symbols $\mathbf{n}$ and $\mathbf{p}$, which mean \emph{nil} and the \emph{current position} (offset from the beginning of the input string).
\medskip

Recall the canonical determinization algorithm that is based on powerset construction:
NFA is simulated on all possible strings,
and the subset of NFA states at each step of the simulation forms a new DFA state,
which is either mapped to an existing identical state or added to the growing set of DFA states.
Since the number of different subsets of NFA states is finite, determinization eventually terminates.
The presence of tags complicates things: it is necessary to track tag values, which depend on the offset that increases at every step.
This makes the usual powerset construction impossible: DFA states augmented with tag values are different and cannot be mapped.
As a result the set of states grows indefinitely and determinization does not terminate.
To address this problem, Laurikari used indirection: instead of storing tag values in TDFA states, he stored value locations --- \emph{registers}.
As long as two TDFA states have the same registers, the actual values in registers do not matter:
they change dynamically at runtime (during TDFA execution), but they do not affect TDFA structure.
A similar approach was used by Grathwohl \cite{Gra15}, who described it as splitting the information contained in a value into static and dynamic parts.
The indirection is not free: it comes at the cost of runtime operations that update register values.
But it solves the termination problem, as the required number of registers is finite, unlike the number of possible register values.
\medskip

From the standpoint of determinization, a TDFA state is a pair.
The first component is a set of configurations $(q, r, l)$ where
$q$ is a TNFA state,
$r$ is a vector of registers (one per tag) and
$l$ is a sequence of tags.
Unlike TNFA simulation that updates tag values immediately when it encounters a tagged transition,
determinization delays the application of tags until the next step.
It records tag sequences along TNFA paths in the $\epsilon$-closure,
but instead of applying them to the current transition,
it stores them in configurations of the new TDFA state
and later applies them to the outgoing transitions.
%
%but the corresponding register operations are attributed to transitions going out of TDFA state, not the incoming one.
This allows filtering tags by the lookahead symbol:
configurations that have no TNFA transitions on the lookahead symbol
do not contribute any register operations to TDFA transition on that symbol.
The use of the lookahead symbol is what distinguishes TDFA(1) from TDFA(0) \cite{Tro17};
it considerably reduces the number of operations and registers.
During $\epsilon$-closure construction configurations are extended to four components $(q, r, h, l)$
where $h$ is the sequence of tags inherited from the origin TDFA state
and $l$ is the new sequence constructed by the $\epsilon$-closure.
\medskip

\begin{algorithm}[] \DontPrintSemicolon \SetKwProg{Fn}{}{}{} \SetAlgoInsideSkip{medskip}
\begin{multicols}{2}
\setstretch{0.9}
\small
\Indm

\nonl\Fn {$\underline{determinization \big( \Sigma, T, Q, q_0, q_f, \Delta \big) } \smallskip$} {
    $S, S_f:$ empty sets of states \;
    $\delta:$ undefined transition function \;
    $\varphi:$ undefined final function \;
    %$\rho:$ precedence function \;
    %$\delta, \varphi, \rho:$ transition, final and precedence functions \;
    $r_0 = \{1, ...\,, |T|\},\; R_f = \{|T|\!+\!1, ...\,, 2|T|\},\; R = \{r_0\} \cup R_f$ \;
    \BlankLine
    $C = epsilon \Xund closure (\{( q_0, r_0, \epsilon, \epsilon )\})$ \;
    $P = precedence (C)$ \;
    $s_0 = add \Xund state (S, S_f, R_f, \varphi, C, P, \epsilon)$ \;
    \BlankLine
    \For {each state $s \in S$} {
        $V:$ map from tag and operation RHS to register \;
        \For {each symbol $a \in \Sigma$} {
            $B = step \Xund on \Xund symbol (s, a)$ \;
            $C = epsilon \Xund closure (B)$ \;
            $O = transition \Xund regops (C, R, V)$ \;
            $P = precedence (C)$ \;
            $s' = add \Xund state (S, S_f, R_f, \varphi, C, P, O)$ \;
            $\delta(s, a) = (s', O)$ \;
        }
    }
    \BlankLine
    \Return TDFA $(\Sigma, T, S, S_f, s_0, R, R_f, \delta, \varphi)$ \;
}
\vspace{2em}

\nonl\Fn {$\underline{add \Xund state \big( S, S_f, R_f, \varphi, C, P, O \big)} \smallskip$} {
    $X = \{ (q, r, l) \mid (q, r, \Xund, l) \in C \}$ \;
    $s = (X, P)$ \;
    \BlankLine
    \If {$s \in S$} {
        \Return $s$
    }
    \BlankLine
    \ElseIf {$\exists s' \in S$ such that $map(s, s', O)$} {
        \Return $s'$ \;
    }
    \BlankLine
    \Else {
        add $s$ to $S$ \;
        \If {$\exists (q, r, l) \in X$ such that $q = q_f$} {
            add $s$ to $S_f$ \;
            $\varphi(s) = final \Xund regops(R_f, r, l)$ \;
        }
        %\BlankLine
        \Return $s$ \;
    }
}
\vspace{2em}

\nonl\Fn {$\underline{map \big( (X, P), (X', P'), O \big)} \smallskip$} {
    \If {$X$ and $X'$ have different subsets of TNFA states \\
            \hphantom{\text{if }} or different lookahead tags for some TNFA state \\
            \hphantom{\text{if }} or precedence is different: $P \neq P'$ } {
        \Return $f\!alse$ \;
    }
    \BlankLine
    $M, M':$ empty maps from register to register \;
    %\BlankLine
    \For {each pair $(q, r, l) \in X$ and $(q, r', l) \in X'$} {
        \For {each $t \in T$} {
            \If {$history(l, t) = \epsilon$ or $t$ is a multi-tag} {
                $i = r[t], \; j = r'[t]$ \;
                %\BlankLine
                \If {both $M[i], M'[j]$ are undefined} {
                    $M[i] = j, \; M'[j] = i$ \;
                    %\BlankLine
                } \ElseIf {$M[i] \neq j$ or $M'[j] \neq i$} {
                    \Return $f\!alse$ \;
                }
            }
        }
    }
    \BlankLine
    \For {each operation $i \leftarrow \Xund$ in $O$} {
        replace register $i$ with $M[i]$ \;
        remove pair $(i, M[i])$ from $M$ \;
    }
    \BlankLine
    \For {each pair $(j, i) \in M$ where $j \neq i$} {
        prepend copy operation $i \leftarrow j$ to $O$ \;
    }
    \BlankLine
    \Return $topological \Xund sort(O)$ \;
}
\vspace{2em}

\nonl\Fn {$\underline{precedence \big( C \big)} \smallskip$} {
    \Return vector $\{q \mid (q, \Xund, \Xund, \Xund)$ in $C \}$ \;
}

%\vfill\null
\columnbreak
\Indp
\SetNlSkip{-0.8em}

\nonl\Fn {$\underline{step \Xund on \Xund symbol \big( (X, P), a \big)} \smallskip$} {
    $B:$ empty sequence of configurations \;
    \For {$(q, r, l) \in X$ ordered by $q$ in the order of $P$} {
        \If {$\exists (q, a, p) \in \Delta \mid a \in \Sigma$} {
            append $(p, r, l, \epsilon)$ to $B$
        }
    }
    \BlankLine
    \Return $B$ \;
}
\vspace{2em}

\nonl\Fn {$\underline{epsilon \Xund closure \big( B \big)} \smallskip$} {
    $C:$ empty sequence of configurations \;
    \BlankLine
    %push configurations in $B$ on stack in reverse order \;
    \For {$(q, r, h, \epsilon)$ in $B$ in reverse order} {
        push $(q, r, h, \epsilon)$ on stack \;
    }
    \BlankLine
    \While {stack is not empty} {
        pop $(q, r, h, l)$ from stack \;
        append $(q, r, h, l)$ to $C$ \;
        \For {each $(q, i, t, p) \in \Delta$ ordered by priority $i$} {
            \If {configuration with state $p$ is not in $C$} {
                push $(p, r, h, lt)$ on stack \;
            }
        }
    }
    \BlankLine
    \Return $\{ (q, r, h, l)$ in $C \mid q = q_f$ or \\
        \hphantom{\Return $\{ (q, r, h, l)$ in $C \mid$} $\exists (q, a, \Xund) \in \Delta$ where $a \in \Sigma \}$  \;
}
\vspace{2em}

\nonl\Fn {$\underline{transition \Xund regops \big( C, R, V \big)} \smallskip$} {
    $O:$ empty list of operations \;
    \For {each $(q, r, h, l) \in C$} {
        \For {each tag $t \in T$} {
            \If {$h_t = history(h, t) \neq \epsilon$ and \;
                    \quad\quad $(history(l, t) = \epsilon$ or $t$ is a multi-tag$)$} {
                $v = regop \Xund rhs(r, h_t, t)$ \;
                $i = V[t][v]$ \;
                \If {$i$ is undefined } {
                    $V[t][v] = i = max \{R\} + 1$ \;
                    $R = R \cup \{i\}$ \;
                }
                \If {operation $i \leftarrow v$ is not in $O$} {
                    append operation $i \leftarrow v$ to $O$ \;
                }
                $r[t] = i$ \;
            }
        }
    }
    \Return $O$ \;
}
\vspace{2em}

\nonl\Fn {$\underline{final \Xund regops \big( R_f, r, l \big)} \smallskip$} {
    $O:$ empty list of operations \;
    \For {each tag $t \in T$} {
        \If {$l_t = history(l, t) \neq \epsilon$} {
            append $R_f[t] \leftarrow regop \Xund rhs(r, l_t, t)$ to $O$ \;
        } \Else {
            append $R_f[t] \leftarrow r[t]$ to $O$ \;
        }
    }
    \Return $O$ \;
}
\vspace{2em}

\nonl\Fn {$\underline{regop \Xund rhs \big( r, h_t, t \big)} \smallskip$} {
    \lIf {$t$ is a multi-tag} {
        \Return $r[t] \cdot h_t$
    } \lElse {
        \Return the last element of $h_t$
    }
}
\vspace{2em}

\nonl\Fn {$\underline{history \big( h, t \big)} \smallskip$} {
    \Switch {$h$} {
        \lCase {$\epsilon$} {
            \Return $\epsilon$
        }
        \lCase {$\;\;\,t \cdot h'$} {
            \Return $\mathbf{p} \cdot history(h')$
        }
        \lCase {$-t \cdot h'$} {
            \Return $\mathbf{n} \cdot history(h')$
        }
        \lCase {$\;\,\,\Xund \cdot h'$} {
            \Return $history(h')$
        }
    }
}

\vfill\null
\end{multicols}
\vspace{1em}
\caption{Determinization of TNFA $(\Sigma, T, Q, q_0, q_f, \Delta)$.
}\label{alg_tdfa}
\end{algorithm}

The second component of TDFA state is precedence information.
It is needed for ambiguity resolution:
if some TNFA state in the $\epsilon$-closure can be reached by different paths, one path must be preferred over the others.
This affects submatch extraction, as the paths may have different tags.
The form of precedence information depends on the disambiguation policy.
We keep the details scoped to functions $precedence$, $step \Xund on \Xund symbol$ and $epsilon \Xund closure$,
so that algorithm \ref{alg_tdfa} can be adapted to different policies without the need to change its structure.
In the case of leftmost greedy policy precedence information is an order on configurations,
represented by $precedence$ as a vector of TNFA states:
$step \Xund on \Xund symbol$ uses it to construct the initial closure,
and $epsilon \Xund closure$ performs depth-first search following transitions from left to right.
POSIX policy is more complex; we do not include pseudocode for it here, but it is extensively covered in \cite{BorTro19}.
\medskip

Algorithm \ref{alg_tdfa} works as follows.
The main function $determinization$ starts by allocating initial registers $r_0$ from $1$ to $|T|$ and final registers $R_f$ from $|T| + 1$ to $2|T|$.
It constructs initial TDFA state $s_0$ as the $\epsilon$-closure of the initial configuration $(q_0, r_0, \epsilon, \epsilon)$.
The initial state $s_0$ is added to the set of states $S$ and the algorithm loops over states in $S$, possibly adding new states on each iteration.
For each state $s$ the algorithm explores outgoing transitions on all alphabet symbols.
Function $step \Xund on \Xund symbol$ follows transitions marked with a given symbol,
and  function $epsilon \Xund closure$ constructs $\epsilon$-closure $C$, recording tag sequences along each fragment of TNFA path.
The set of configurations in the $\epsilon$-closure forms a new TDFA state $s'$.
Function $transition \Xund regops$ uses the $h$-components of configurations in $C$ to construct register operations on transition from $s$ to $s'$.
The same register is allocated for all outgoing transitions with identical operation right-hand-sides,
but different tags do not share registers,
and vacant registers from other TDFA states are not reused
(these rules ensure that there are no artificial dependencies between registers, which makes optimizations easier without the need to construct SSA).
The new state $s'$ is inserted into the set of states $S$:
function $add \Xund state$ first tries to find an identical state in $S$;
if that fails, it looks for a state that can be mapped to $s'$;
if that also fails, $s'$ is added to $S$.
If the new state contains the final TNFA state, it is added to $S_f$,
and the $final \Xund regops$ function constructs register operations for the final \emph{quasi-transition}
which does not consume input characters and gets executed only once at the end of match.
\medskip

TDFA states are considered identical if both components (configuration set and precedence) coincide.
States that are not identical but differ only in registers can be made identical (mapped), provided that there is a bijection between registers.
Function $map$ attempts to construct such a bijection $M$:
for every tag, and for each pair of configurations
it adds the corresponding pair of registers to $M$.
If either of the two registers is already mapped to some other register, bijection cannot be constructed.
For single-valued tags mapping ignores configurations that have the tag in the lookahead sequence ---
every transition out of TDFA state overwrites tag value with a set operation, making the current register values obsolete.
For multi-valued tags this optimization is not possible, because append operations do not overwrite previous values.
If the mapping has been constructed successfully, $map$ updates register operations:
for each pair of registers in $M$ it adds a copy operation,
unless the left-hand-side is already updated by a set or append operation,
in which case it replaces left-hand-side with the register it is mapped to.
The operations are topologically sorted ($topological \Xund sort$ is defined on page \pageref{alg_opt2});
in the presence of copy and append operations this is necessary to ensure that old register values are used before they are updated.
Topological sort ignores trivial cycles such as append operation $i \leftarrow i \cdot h$,
but if there are nontrivial cycles the mapping is rejected
(handling such cycles requires a temporary register, which makes control flow more complex and inhibits optimizations).
\medskip

After determinization is done, the information in TDFA states is erased --- it is no longer needed for TDFA execution.
States are just atomic values with no internal structure.
Disambiguation decisions are embedded in TDFA; there is no disambiguation at runtime.
The only runtime overhead compared to an ordinary DFA is the execution of register operations on transitions.
A TDFA may have more states than a DFA for the same RE with all tags removed, because states that can be mapped in a DFA cannot always be mapped in a TDFA.
Minimization can reduce the number of states,
especially if it is applied after register optimizations that can get rid of many operations and make more states compatible.
We focus on optimizations in section \ref{section_impl}.
%The difference between TDFA and an ordinary DFA is the presence of registers and register operations on transitions.
\medskip

\begin{figure}%[t!]
\includegraphics[width=\linewidth]{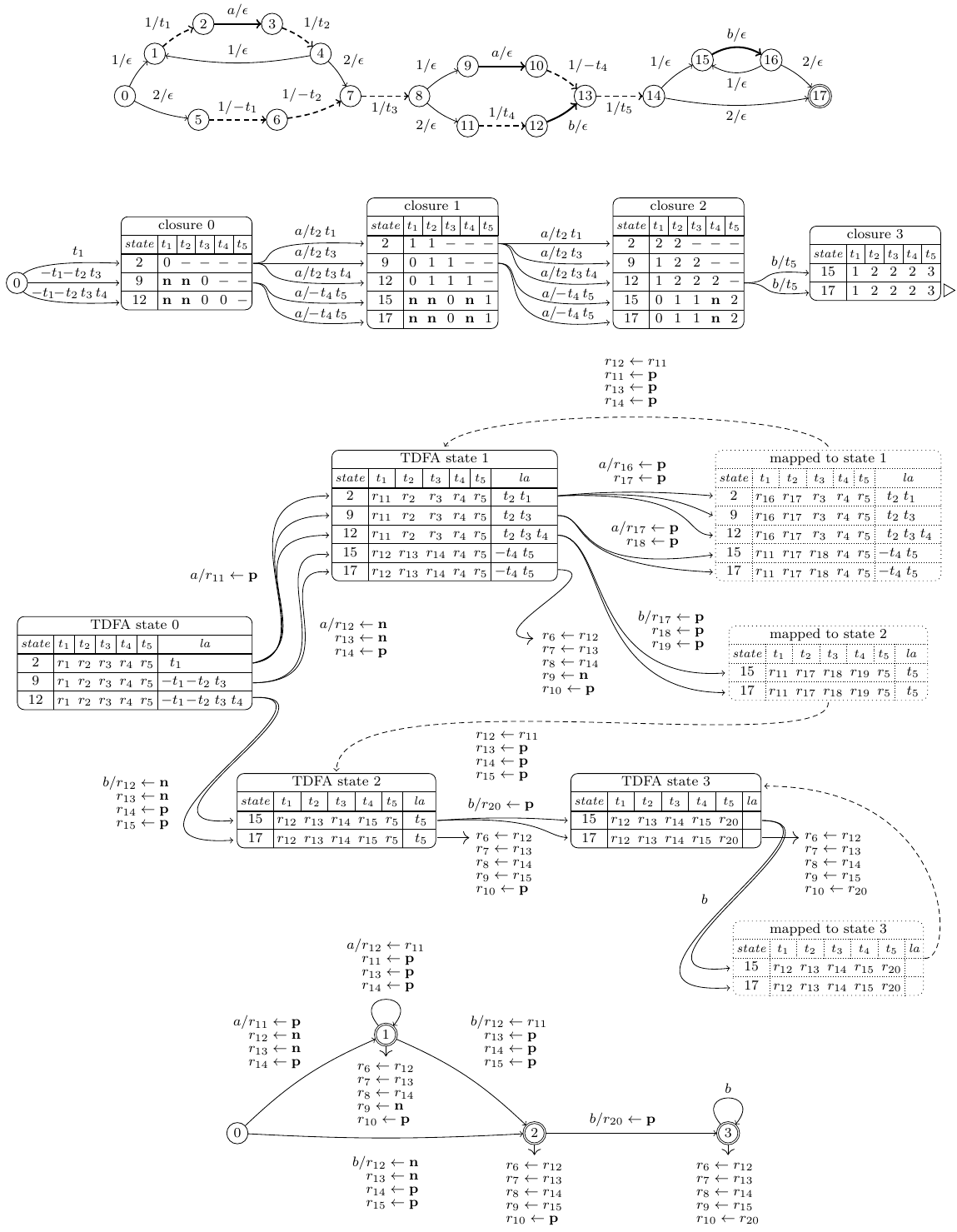}
\vspace{0.1em}
\caption{
Example for a RE $(1a2)^*3(a|4b)5b^*$: TNFA, simulation on string $aab$, determinization, TDFA.
%TDFA construction for RE $(1a2)^*3(a|4b)5b^*$\\
%(top: TNFA, middle: determinization process, bottom: the resulting TDFA with final registers $r_6$ to $r_{10}$).
}\label{fig:tdfa}
\end{figure}

Figure \ref{fig:tdfa} shows an example of TDFA construction:

\medskip

\begin{itemize}

\item[\ding{212}]
The RE is $(1a2)^*3(a|4b)5b^*$. It defines language $\{a^n b^m \mid n + m > 0 \}$
and has five tags $t_1, t_2, t_3, t_4, t_5$.
\medskip

\item[\ding{212}]
TNFA has three kinds of transitions:
bold transitions on alphabet symbols (four of them for each symbol in the RE),
thin $\epsilon$-transitions with priority
and dashed $\epsilon$-transitions with priority and tag.
Tags $t_1$, $t_2$ are under repetition, so the zero-repetition path $0 \!\rightarrow\! 5 \!\rightarrow\! 6 \!\rightarrow\! 7$ contains transitions with negative tags $-t_1$, $-t_2$.
Likewise tag $t_4$ is in alternative, so path $8 \!\rightarrow\! 9 \!\rightarrow\! 10 \!\rightarrow\! 13$ contains transition with negative tag $-t_4$.
Tags $t_3$, $t_5$ are in top-level concatenation and do not need negative tags.
\medskip

\item[\ding{212}]
TNFA simulation on string $aab$ consists of four steps.
The first step starts with state $0$.
Every other step starts with states from the previous step and follows transitions labeled with the current symbol.
Each step constructs $\epsilon$-closure by following $\epsilon$-paths and collecting tag sequence along the way.
The value of positive tags in the corresponding row of the closure matrix is set to the number of characters consumed so far.
The value of negative tags is set to nil $\mathbf{n}$.
The value of tags not in the sequence is inherited from the previous closure.
Simulation ends when all characters have been consumed.
Since the last closure contains a row with the final state $17$, it is a match and the final tag values are $1, 2, 2, 2, 3$.
\medskip

\item[\ding{212}]
The match is ambiguous: it is possible to match $aab$ by following path $0 \!\rightsquigarrow\! 2 \!\rightsquigarrow\! 2 \!\rightsquigarrow\! 12 \!\rightsquigarrow\! 17$
(let the greedy repetition consume $aa$)
but it is also possible to follow path $0 \!\rightsquigarrow\! 2 \!\rightsquigarrow\! 9 \!\rightsquigarrow\! 15 \!\rightsquigarrow\! 17$
(let the greedy repetition consume only the first $a$).
The first match is preferable by both POSIX and leftmost-greedy policies.
\medskip

\item[\ding{212}]
Determinization is similar to simulation, but TDFA states store registers instead of offsets.
This solves the problem of mapping states that differ only in tag values:
for example, closures $1$ and $2$ cannot be mapped, although they have identical states and tag sequences,
but TDFA state corresponding to closure $2$ is mapped to state $1$.
This is possible due to the register operations on the dashed backward transition.
Note that there is one copy operation $r_{12} \leftarrow r_{11}$,
but other copy operations for $r_{11}$, $r_{13}$, $r_{14}$ are combined with set operations,
e.g. $r_{11} \leftarrow \mathbf{p}$ is the combination of $r_{16} \leftarrow \mathbf{p}$ and $r_{11} \leftarrow r_{16}$
(see lines 43 -- 45 of algorithm \ref{alg_tdfa}).
\medskip

\item[\ding{212}]
Unlike simulation, determinization does not immediately apply tag sequences to registers.
Instead, it stores them as part of TDFA state (in the \emph{lookahead} column, shortened as \emph{la}).
Compare tag sequences on transitions to closures $0$, $1$, $2$ to that in states $0$, $1$, $2$ respectively ---
these are the same tags.
Lookahead tags form register operations on the outgoing transitions:
e.g. lookahead tag $t_1$ in the first row of TDFA state $0$
(corresponding to tagged TNFA transition $1 \!\rightarrow\! 2$)
forms operation $r_{11} \leftarrow \mathbf{p}$.
\medskip

\item[\ding{212}]
For every distinct set or append operation $transition \Xund regops$ allocates a new register and stores the updated tag value in it.
Note that it would be impossible to reuse the same register
(e.g. to have $r_{1} \leftarrow \mathbf{p}$ instead of $r_{11} \leftarrow \mathbf{p}$ on transition from state $0$ to $1$)
because there may be conflicting operations
(e.g. $r_{12} \leftarrow \mathbf{n}$ for lookahead tag $-t_1$).
Therefore tag $t_1$ in TDFA state $1$ is represented with two registers $r_{11}$ and $r_{12}$,
reflecting the fact that state $1$ may be reached by different TNFA paths with conflicting submatch values.
\medskip

\item[\ding{212}]
Final TDFA states are all states containing TNFA state $17$ (i.e. states $1$, $2$ and $3$).
In addition to normal transitions final TDFA states have quasi-transitions
that set final registers $r_6$ -- $r_{10}$.
These quasi-transitions do not consume any symbol, and the operations on them are executed once at the end of match.
\medskip

\item[\ding{212}]
In the resulting TDFA all internal structure in the states is erased,
leaving atomic states with transitions and register operations.
Registers can be renamed to occupy consecutive numbers,
and the number of registers and operations can be reduced (see section \ref{section_impl}).
\medskip

\end{itemize}

\FloatBarrier

\section{Implementation}\label{section_impl}

In this section we describe optimizations and practical details that should be taken into account when implementing TDFA.
None of the optimizations is particularly complex or vital for TDFA operation,
but applied together and in the correct order they can make TDFA considerably faster and smaller.

\subsection{Multi-valued tags}

The most straightforward representation of multi-valued tags is a vector of offsets.
It is very inefficient because copy and append operations need to copy entire vectors (which could grow arbitrarily long).
A more efficient representation is a \emph{prefix tree}.
It is possible because tag sequences in the operations map on the path tree constructed by the $\epsilon$-closure.
The tree can be stored as an array of nodes $(pred, o\!f\!\!f\!s)$ where
$pred$ is the index of a predecessor node, and $o\!f\!\!f\!s$ is a positive or negative tag value.
Individual sequences in the tree are addressed by integer indices of tree nodes
(zero index corresponds to the empty sequence).
This representation is space efficient (common prefixes are shared),
but most importantly it makes copy operations as simple as copying scalar values (tree indices).
Append operations are more difficult, as they require a new slot (or a couple of slots) in the prefix tree.
However, if the backing array is allocated in large chunks of memory,
then the amortized complexity of each operation is constant.
This representation was used by multiple researches,
e.g. Karper describes it as the \emph{flyweight pattern} \cite{Kar14}.

\subsection{Fallback operations}

In practice it is often necessary to match the longest possible prefix of a string rather than the whole string.
After matching a short prefix, TDFA may attempt to match a longer prefix.
If that fails, it must fallback to the previous final state and restore the input position accordingly.
A final state is also a \emph{fallback state} if there are non-accepting paths out of it,
and a path is non-accepting if does not go through another final state
(which may happen either because the input characters do not match or due to a premature end of input).
\medskip

For an ordinary DFA the only information that should be saved in a fallback state is the input position.
For TDFA it is also necessary to backup registers that may be clobbered on the non-accepting paths from the fallback state.
Backup operations should be added on transitions out of the fallback state,
and restore operations should be added on the fallback quasi-transition,
which replaces the final quasi-transition for fallback paths.
Final registers can be reused for backups, as by TDFA construction they are used only on the final quasi-transitions.
Backup registers are only needed for copy and append operations (set operations do not depend on registers).
\medskip

\begin{algorithm}[] \DontPrintSemicolon \SetKwProg{Fn}{}{}{} \SetAlgoInsideSkip{medskip}
\setstretch{0.9}
\small

\begin{multicols}{2}
\Indm

\nonl\Fn {$\underline{f\!allback \Xund regops \big( \big)} \smallskip$} {
    $\psi:$ undefined fallback function \;
    \BlankLine
    \For {each fallback state $s \in S$} {
        $O:$ empty list of register operations \;
        \BlankLine
        \For {each operation on quasi-transition $\varphi(s)$} {
            \If {append $i \leftarrow j \cdot h$ and $j$ is clobbered} {
                $backup \Xund regops(s, i, j)$ \;
                append operation $i \leftarrow i \cdot h$ to $O$
            }
            \ElseIf {copy $i \leftarrow j$ and $j$ is clobbered} {
                $backup \Xund regops(s, i, j)$
            }
            \Else {
                append a copy of this operation to $O$
            }
        }
        \BlankLine
        $\psi(s) = O$
    }
    \BlankLine
    \Return $\psi$ \;
}

\columnbreak

\nonl\Fn {$\underline{backup \Xund regops \big( s, i, j \big)} \smallskip$} {
    \For {each alphabet symbol $a \in \Sigma$} {
        $(s', O) = \delta(s, a)$ \;
        \If {exist non-accepting paths from $s'$} {
            append copy operation $i \leftarrow j$ to $O$
        }
    }
}

\vfill\null
\end{multicols}

\vspace{1em}
\caption{Adding fallback operations to TDFA
$(\Sigma, T, S, S_f, s_0, R, R_f, \delta, \varphi)$.
}\label{alg_fallback}
\end{algorithm}
%\vspace{2em}

Algorithm \ref{alg_fallback} shows how to add such operations.
It assumes that fallback states and clobbered registers for each fallback state have already been identified.
This can be done as follows.
First, augment TDFA with a \emph{default state} that makes transition function $\delta$ total
(if a premature end of input is possible, add a quasi-transition from non-final states to the default state).
Then compute reachability of the default state by doing backward propagation from states that have transitions to it.
If the default state is reachable from a final state, then it is a fallback state.
Clobbered registers can be found by doing depth-first search from a fallback state,
visiting states from which the default state is reachable,
and accumulating left-hand-sides of register operations.

\subsection{Register optimizations}

TDFA induces a \emph{control flow graph} (CFG) with three kinds of nodes:
\medskip

\begin{itemize}
\item[$\bullet$] \emph{basic blocks} for register operations on symbolic transitions
\item[$\bullet$] \emph{final blocks} for final register operations
\item[$\bullet$] \emph{fallback blocks} for fallback register operations
\end{itemize}
\medskip

\begin{figure}[]
\includegraphics[width=\linewidth]{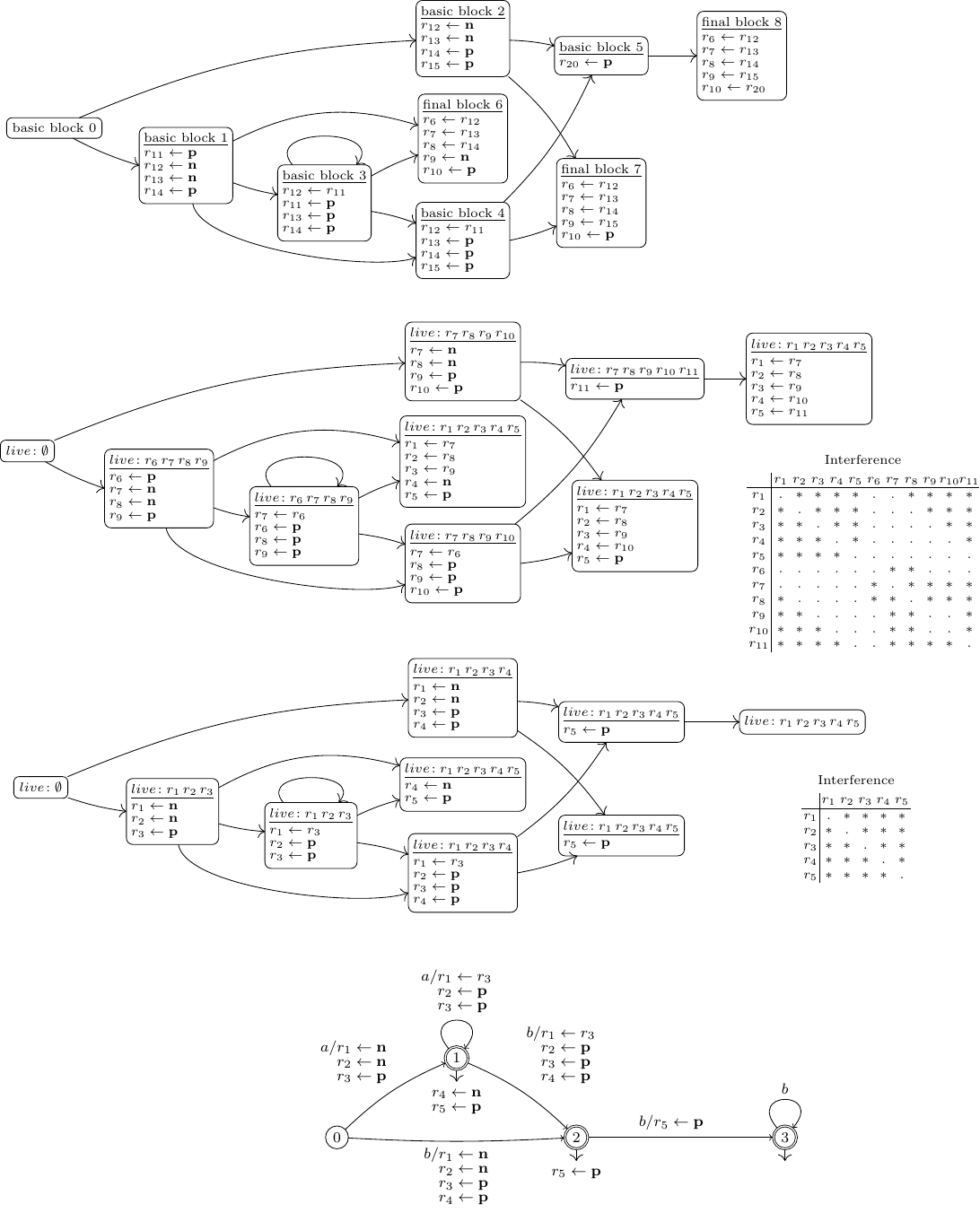}
\vspace{0.5em}
\caption{
Register optimizations for TDFA on figure \ref{fig:tdfa}. \\
Top to bottom: initial CFG,
CFG after compaction with per-block liveness information and interference table,\\
CFG on the second round of optimizations,
optimized TDFA with final registers $r_1$ to $r_5$.
}\label{fig:tdfa_regopt}
\end{figure}

There is an arc between two blocks in CFG if one is reachable from another in TDFA without passing through other register operations.
Additionally, fallback blocks have arcs to all blocks reachable by TDFA paths that may fall through to these blocks.
Figure \ref{fig:tdfa_regopt} shows CFG for the example from section \ref{section_tdfa}.
\medskip

CFG represents a program on registers, so the usual compiler optimizations can be applied to it,
resulting in significant reduction of registers and operations.
RE2C uses the following optimization passes for the number of repetitions $N=2$
(pseudocode is given by the algorithms \ref{alg_opt1} and \ref{alg_opt2}):
\medskip

\begin{enumerate}
    \item Compaction
    \item Repeat $N$ times:
    \begin{enumerate}[label=\alph*.]
        \item Liveness analysis
        \item Dead code elimination
        \item Interference analysis
        \item Register allocation with copy coalescing
        \item Local normalization
    \end{enumerate}
\end{enumerate}
\medskip

%\pagebreak

\begin{algorithm}[] \DontPrintSemicolon \SetKwProg{Fn}{}{}{} \SetAlgoInsideSkip{medskip}
\begin{multicols}{2}
\setstretch{0.9}
\small

\Indm

\nonl\Fn {$\underline{optimizations \big( G \big)} \smallskip$} {
    $V = compaction(G)$ \;
    $G = renaming(G, V)$ \;
    \For {$i = \overline{1,2}$} {
        $L = liveness \Xund analysis(G)$ \;
        $dead \Xund code \Xund elimination(G, L)$ \;
        $I = inter\!f\!erence \Xund analysis(G, L)$ \;
        $V = register \Xund allocation (G, I)$ \;
        $renaming(G, V)$ \;
        $normalization(G)$ \;
    }
}
\vspace{2em}

\nonl\Fn {$\underline{renaming \big( G, V \big)} \smallskip$} {
    \For {each block $b$ in $G$} {
        \For {each operation in $b$} {
            \If {set operation $i \leftarrow v$} {
                rename $i$ to $V[i]$ \;
            }
            \If {copy or append operation $i \leftarrow j ...$} {
                rename $i$ to $V[i]$ and $j$ to $V[j]$ \;
            }
        }
    }
}
\vspace{2em}

\nonl\Fn {$\underline{liveness \Xund analysis \big( G \big)} \smallskip$} {
    $L:$ boolean matrix indexed by blocks and registers \;
    \BlankLine
    \For {each block $b$ in $G$} {
        \For {each register $i$ in $G$} {
            $L[b][i] = f\!alse$ \;
        }
    }
    \BlankLine
    \For {each final block $b$ in $G$} {
        \For {each final register $i$ in $G$} {
            $L[b][i] = true$ \;
        }
    }
    \BlankLine
    \While {$true$} {
        $fix = true$ \;
        \For {each basic block $b$ in $G$ in post-order} {
            $L_b = $ copy of row $L[b]$ \;
            \For {each successor $s$ of block $b$} {
                $L_s = $ copy of row $L[s]$ \;
                \For {each operation in $s$ in post-order} {
                    \If {set operation $i \leftarrow v$} {
                        $L_s[i] = f\!alse$ \;
                    }
                    \If {copy operation $i \leftarrow j$} {
                        \If {$L_s[i]$} {
                            $L_s[i] = f\!alse$ \;
                            $L_s[j] = true$ \;
                        }
                    }
                }
                \BlankLine
                \For {each register $i$ in $G$} {
                    $L_b[i] = L_b[i] \vee L_s[i]$
                }
            }
            \BlankLine
            \If {$L[b] \neq L_b$} {
                $L[b] = L_b$ \;
                $fix = f\!alse$ \;
            }
        }
        \lIf {$fix$} {$break$}
    }
    \BlankLine
    \For {each fallback block $b$ in $G$} {
        \For {each final register $i$ in $G$} {
            $L[b][i] = true$ \;
        }
        \BlankLine
        $L_b = $ copy of row $L[b]$ \;
        \For {each operation $i \leftarrow \Xund$ in $b$} {
            $L_b[i] = f\!alse$ \;
        }
        \For {each copy or append operation $\Xund \leftarrow j ...$ in $b$} {
            $L_b[j] = true$ \;
        }
        \BlankLine
        \For {each block $s$ in $G$ that may fall through to $b$} {
            \For {each register $i$ in $G$} {
                $L[s][i] = L[s][i]$ or $L_b[i]$ \;
            }
        }
    }
    \BlankLine
    \Return $L$ \;
}

\vfill\null

\columnbreak

\Indp
\SetNlSkip{-0.8em}

\nonl\Fn {$\underline{compaction \big( G \big)} \smallskip$} {
    $U:$ boolean vector indexed by registers \;
    $V:$ integer vector indexed by registers \;
    \BlankLine
    \For {each register $i$ in $G$} {
        $U[i] = f\!alse$ \;
    }
    \For {each block $b$ in $G$} {
        \For {each operation in $b$} {
            \If {set operation $i \leftarrow v$} {
                $U[i] = true$ \;
            }
            \If {copy or append operation $i \leftarrow j ...$} {
                $U[i] = U[j] = true$ \;
            }
        }
    }
    $n = 0$ \;
    \For {registers $i$ in $G$ such that $U[i]$} {
        $n = n + 1, \; V[i] = n$ \;
    }
    \BlankLine
    \Return $V$ \;
}
\vspace{2em}

\nonl\Fn {$\underline{dead \Xund code \Xund elimination \big( G, L \big)} \smallskip$} {
    \For {each basic block $b$ in $G$} {
        $L_b = $ copy of row $L[b]$ \;
        \For {each operation $i \leftarrow \Xund$ in $b$ in post-order} {
            \If {$L_b[i]$} {
                \If {set operation $i \leftarrow v$} {
                    $L_b[i] = f\!alse$ \;
                }
                \If {copy operation $i \leftarrow j$} {
                    $L_b[i] = f\!alse$ \;
                    $L_b[j] = true$ \;
                }
            } \lElse {
                remove dead operation
            }
        }
    }
}
\vspace{2em}

\nonl\Fn {$\underline{inter\!f\!erence \Xund analysis \big( G, L \big)} \smallskip$} {
    $I:$ boolean matrix indexed by registers \;
    $V:$ vector of histories indexed by registers \;
    \BlankLine
    \For {each register $i$ in $G$} {
        \For {each register $j$ in $G$} {
            $I[i][j] = I[j][i] = f\!alse$ \;
        }
    }
    \BlankLine
    \For {each block $b$ in $G$} {
        \For {each copy or append operation $i \leftarrow j ...$ in $b$} {
            $V[j] = j$ \;
        }
        \BlankLine
        \For {each operation in $b$} {
            $I_b = $ copy of row $L[b]$ \;
            \BlankLine
            \If {set operation $i \leftarrow v$} {
                $V[i] = v$ \;
                $I_b[i] = f\!alse$ \;
            } \ElseIf {copy operation $i \leftarrow j$} {
                $V[i] = V[j]$ \;
                $I_b[i] = I_b[j] = f\!alse$ \;
            } \ElseIf {append operation $i \leftarrow j \cdot h$} {
                $V[i] = V[j] \cdot h$ \;
            }
            \BlankLine
            \For {operations $k \leftarrow \Xund$ in $b$ with $V[k] = V[i]$} {
                $I_b[k] = f\!alse$ \;
            }
            \BlankLine
            \For {registers $k$ in $G$ such that $I_b[k]$} {
                $I[i][k] = I[k][i] = true$ \;
            }
        }
    }
    \BlankLine
    \For {registers $i$ in $G$ not used in append operations} {
        \For {registers $j$ in $G$ used in append operations} {
            $I[i][j] = I[j][i] = true$ \;
        }
    }
    \BlankLine
    \Return $I$ \;
}

\vfill\null

\end{multicols}
\vspace{1em}
\caption{Register optimizations (part 1).}\label{alg_opt1}
\end{algorithm}

\begin{algorithm}[t] \DontPrintSemicolon \SetKwProg{Fn}{}{}{} \SetAlgoInsideSkip{medskip}
\begin{multicols}{2}
\setstretch{0.9}
\small

\Indm

\nonl\Fn {$\underline{register \Xund allocation \big( G, I \big)} \smallskip$} {
    $V:$ vector of registers indexed by registers \;
    $B:$ vector of registers indexed by registers \;
    $S:$ vector of register sets indexed by registers \;
    \BlankLine
    \For {each register $i$ in $G$} {
        $B[i] = -1$ \;
        $S[i] = \emptyset$ \;
    }
    \BlankLine
    \For {each block $b$ in $G$} {
        \For {each operation in $b$} {
            \If {copy or append $i \leftarrow j ...$ and $i \neq j$} {
                $x = B[i], \; y = B[j]$ \;
                \If {$x = -1$ and $y = -1$ and $\neg I[i][j]$} {
                    $B[i] = B[j] = i$ \;
                    $S[i] = \{i, j\}$ \;
                } \ElseIf {$x \neq -1$ and $y = -1$} {
                    \If {$\forall k \in S[x]: \neg I[k][j]$} {
                        $B[j] = x$ \;
                        $S[x] = S[x] \cup \{j\}$ \;
                    }
                } \ElseIf {$x = -1$ and $y \neq -1$} {
                    \If {$\forall k \in S[y]: \neg I[k][i]$} {
                        $B[i] = y$ \;
                        $S[y] = S[y] \cup \{i\}$ \;
                    }
                }
            }
        }
    }
    \BlankLine
    \For {registers $i$ in $G$ such that $B[i] = i$} {
        \For {registers $j$ in $G$ such that $B[j] = j$ and $j > i$} {
            \If {$\forall i \in S[x], j \in S[y]: \neg I[i][j]$} {
                $B[y] = x$ \;
                $S[x] = S[x] \cup S[y]$ \;
                $S[y] = \emptyset$ \;
            }
        }
    }
    \BlankLine
    \For {registers $i$ in $G$ such that $B[i] = -1$} {
        \If {$\exists j$ in $G: B[j] = j$ and $\forall k \in S[j]: \neg I[i][k]$} {
            $B[i] = j$ \;
            $S[j] = S[j] \cup \{i\}$ \;
        } \Else {
            $B[i] = i$ \;
            $S[i] = \{i\}$ \;
        }
    }
    \BlankLine
    $n = 0$ \;
    \For {registers $i$ in $G$ such that $B[i] = i$} {
        $n = n + 1$ \;
        \For {registers $j \in S[i]$} {
            $V[j] = n$ \;
        }
    }
    \BlankLine
    \Return $V$ \;
}

\vfill\null

\columnbreak

\nonl\Fn {$\underline{normalization \big( G \big)} \smallskip$} {
    \For {each block $b$ in $G$} {
        \For {each contiguous set operation range $O$} {
            $remove \Xund duplicates(O)$ \;
            $sort(O)$ \;
        }
        \For {each contiguous copy operation range $O$} {
            $remove \Xund duplicates(O)$ \;
            $topological \Xund sort(O)$ \;
        }
        \For {each contiguous append operation range $O$} {
            $remove \Xund duplicates(O)$ \;
        }
    }
}
\vspace{2em}

\nonl\Fn {$\underline{topological \Xund sort \big( O \big)} \smallskip$} {
    $I:$ vector of in-degree indexed by registers \;
    \For {each copy or append operation $i \leftarrow j ...$ in $O$} {
        $I[i] = I[j] = 0$ \;
    }
    \For {each copy or append operation $\Xund \leftarrow j ...$ in $O$} {
        $I[j] = I[j] + 1$ \;
    }
    \BlankLine
    $O':$ empty list of operations \;
    $nontrivial \Xund cycle = f\!alse$ \;
    \BlankLine
    \While {$O$ is not empty} {
        \For {each operation $i \leftarrow \Xund$ in $O$} {
            \If {$I[i] = 0$} {
                remove operation from $O$, append to $O'$ \;
                \If {this is a copy/append operation $i \leftarrow j ...$} {
                    $I[j] = I[j] - 1$ \;
                }
            }
        }
        \If {nothing added to $O'$ but $O$ is not empty} {
            \If {$\exists$ operation $i \leftarrow j ...$ in $O$ with $i \neq j$} {
                $nontrivial \Xund cycle = true$
            }
            append $O$ to $O'$ \;
            $break$ (only cycles left)
        }
    }
    \BlankLine
    $O = O'$ \;
    \Return $\not nontrivial \Xund cycle$ \;
}
\vspace{2em}

\nonl\Fn {$\underline{remove \Xund duplicates \big( O \big)} \smallskip$} {
    \For {each operation $o$ in $O$} {
        \For {each subsequent operation $o' = o$ in $O$} {
            remove duplicate operation $o'$
        }
    }
}

\vfill\null

\end{multicols}
\vspace{0.5em}
\caption{Register optimizations (part 2).}\label{alg_opt2}
\end{algorithm}

Compaction pass is applied only once immediately after determinization.
It renames registers so that they occupy contiguous range of numbers with no ``holes''.
This is needed primarily to allow other optimization passes use registers as indices in liveness and interference matrices.
\medskip

Liveness analysis builds a boolean 2-dimensional matrix indexed by CFG blocks and registers.
A cell $L[b][i]$ is true iff register $i$ is alive in block $b$ (meaning that its value is used).
Function $liveness \Xund analysis$ uses iterative data-flow approach.
Initially only the final registers in the final blocks are alive.
The algorithm iterates over CFG blocks in post-order, expanding the live set, until it reaches a fix point.
Lastly it marks backup registers as alive in all blocks reachable from fallback blocks by non-accepting paths.
\medskip

Dead code elimination removes operations whose left-hand-side register is not alive.
\medskip

Interference analysis builds a boolean square matrix indexed by registers.
A cell $I[i][j]$ is true iff registers $i$ and $j$ interfere with each other
(their lifetimes overlap, so they cannot be represented with one register).
Initially none of the registers interfere.
Function $inter\!f\!erence \Xund analysis$ considers each CFG block $b$ and inspects each register $j$ used on the right-hand-side of an operation:
all registers alive in block $b$ interfere with $j$,
except for registers that have the same value (tracked by the vector $V$).
Finally registers for multi-valued and single-valued tags are marked as interfering with each other.
\medskip

Register allocation partitions registers into equivalence classes.
Registers inside of one class do not interfere with each other, so they can all be replaced with a single \emph{representative}.
Initially none of the registers belongs to any class.
Function $register \Xund allocation$ loops over copy operations
and tries to put source and destination into one class (so that the copy can be removed).
Vector $B$ maps registers to their representative, and vector $S$ maps representatives to their class.
The algorithm tries to merge non-interfering equivalence classes,
and then puts the remaining registers into an non-interfering class (allocating a new class if necessary).
Finally it maps representatives to consecutive numbers and stores them in the $V$ vector.
The constructed partitioning is not minimal, but it's a good approximation,
since finding the minimal clique cover of a graph is NP-complete.
\medskip

Local normalization reconciles operations after previous passes.
The $normalization$ function removes duplicate operations that might appear after different registers are collapsed into one.
It also sorts operations, so that operation sequences could be compared easily
(which is used in further optimizations like in minimization).
Each continuous range of set, copy or append operations is handled separately,
because operations of different kinds should not be reordered (that could change the end result).
\medskip

Figure \ref{fig:tdfa_regopt} is a continuation of example on figure \ref{fig:tdfa}:
\medskip

\begin{itemize}

\item[\ding{212}]
The CFG contains $9$ blocks: basic blocks $0$ -- $5$ (one for each tagged transition on symbol in TDFA, plus the start block $0$)
and final blocks $6$ -- $8$ (one per each final quasi-transition).
There are no fallback blocks in this example, because there are no fallback states in TDFA:
every transition out of a final state goes to another final state,
so the attempt to match a longer string will either fail immediately (before leaving the final state),
or it will succeed immediately.
\medskip

\item[\ding{212}]
The second CFG is after compaction and the first pass of liveness and interference analysis.
Compaction renames registers $r_6$ -- $r_{15}$ and $r_{20}$ to $r_1$ -- $r_{11}$,
reducing the size of the register range from 20 to 11 and the size of the liveness and interference matrices almost 2x and 4x respectively.
Liveness information is shown at the top of each block.
Interference matrix uses asterisk for interfering register pairs and dot for non-interfering ones.
It can be seen that there are many dots in the table, which means optimization opportunities.
The interference matrix is symmetrical, as the interference relation is commutative.
\medskip

\item[\ding{212}]
The third CFG is after the second pass of liveness and interference analysis.
The number of registers is reduced from $11$ to $5$.
Many operations have been eliminated, for example the copy operation $r_1 \leftarrow r_7$ in the final block $8$ of the second CFG
was removed by copy coalescing, because registers $1$ and $7$ did not interfere and register allocation put them in one equivalence class
(and likewise for the other copy operations in block $8$).
In basic blocks $1$ and $3$ set operations $r_6 \leftarrow \mathbf{p}$ and $r_9 \leftarrow \mathbf{p}$ were collapsed into $r_3 \leftarrow \mathbf{p}$
after non-interfering registers $r_6$ and $r_9$ had been renamed to $r_3$.
Interference matrix has dots only on the main diagonal (a register does not interfere with itself),
which leaves no room for further optimization.
\medskip

\item[\ding{212}]
The resulting optimized TDFA is at the bottom of figure \ref{fig:tdfa_regopt}.
The final registers are now $r_1$ -- $r_5$.
\medskip

\end{itemize}

\subsection{Minimization}

Minimization can be applied to TDFA in the same way as to an ordinary DFA (e.g. the Moore's algorithm),
except that transitions on the same alphabet symbol but with different register operations
should be treated as different transitions, so their destination states cannot be merged.
To get optimal performance minimization algorithm should be able to compare operations on transitions in constant time.
This is possible if operation sequences are inserted into a hash map and represented with unique numeric identifiers.
Such a comparison may have false negatives, as non-identical operations lists may be equivalent
(e.g. $r_1 \leftarrow \mathbf{p}, r_2 \leftarrow \mathbf{n}$ is not identical, but equivalent to $r_2 \leftarrow \mathbf{n}, r_1 \leftarrow \mathbf{p}$).
False negatives do not affect minimization correctness, but the end result may be suboptimal.
To avoid that, minimization should be applied after register optimizations (which may remove some register operations)
and most importantly after normalization (defined on page \pageref{alg_opt2}).

\subsection{Fixed tags}

\newcommand\nan{N\!a\!N}
\newcommand\nobasetag{-1}%{N\!otag}

\emph{Fixed tags} is a very important optimization that happens at the RE level.
In cases with high tag density (such as POSIX REs with nested submatch groups)
this optimization alone may be more effective than all register optimizations combined.
The key observation is,
if a pair of tags is within fixed distance from each other, there is no need to track both of them:
the value of one tag can be computed from the value of the other tag one by adding a fixed offset.
This optimization is fast (linear in the size of a RE)
and has the potential to reduce both TDFA construction time and matching time.
\medskip

Algorithm \ref{alg_fixed_tags} finds fixed tags by performing top-down structural recursion on a RE.
It has four parameters: $e$ is the current sub-RE, $t$ is the current base tag,
$d$ is the distance to base tag, and $k$ is the distance to the start of the current level.
Levels are parts of a RE where any two points either both match or both do not match.
A level increases on recursive descent into alternative or repetition subexpressions, but not concatenation.
Tags on different levels should not be fixed on each other, even if they are within fixed distance on any path that goes through both of them,
because there are paths that go through only one tag (so the other one is nil).
Tag value $-1$ denotes the absence of base tag: when descending to the next level initially there is no base tag, and the first tag on the current level becomes the base.
One exception is the top level, where the initial base tag should be a special value denoting the rightmost position (which is always known at the end of the match).
The algorithm recursively returns the new base tag, the updated distance to base tag, and the updated level distance.
Special distance value $\nan$ (not-a-number) is understood to be a fixed point in arithmetic expressions: any expression involving $\nan$ amounts to $\nan$.

\begin{multicols}{2}
\begin{algorithm}[H] \DontPrintSemicolon \SetKwProg{Fn}{}{}{} \SetAlgoInsideSkip{medskip}
\setstretch{0.9}
\small

\Indm

\nonl\Fn {$\underline{fixed \Xund tags \big( e, t, d, k \big)} \smallskip$} {
    \If {$e = \epsilon$} {
        \Return $t, d, k$
    }
    \BlankLine
    \ElseIf {$e = a \in \Sigma$} {
        \Return $t, d + 1, k + 1$
    }
    \BlankLine
    \ElseIf {$e = e_1 | e_2$} {
        $\Xund, \Xund, k_1 = fixed \Xund tags(e_1, \nobasetag, \nan, 0)$ \;
        $\Xund, \Xund, k_2 = fixed \Xund tags(e_2, \nobasetag, \nan, 0)$ \;
        \If {$k_1 = k_2$} {
            \Return $t, d + k_1, k + k_1$
        }
        \Return $t, \nan, \nan$
    }
    \BlankLine
    \ElseIf {$e = e_1 e_2$} {
        $t_2, d_2, k_2 = fixed \Xund tags(e_2, t, d, k)$ \;
        $t_1, d_1, k_1 = fixed \Xund tags(e_1, t_2, d_2, k_2)$ \;
        \Return $t_1, d_1, k_1$
    }
    \BlankLine
    \ElseIf {$e = e_1^{n, m}$} {
        $\Xund, \Xund, k_1 = fixed \Xund tags(e_1, \nobasetag, \nan, 0)$ \;
        \If {$n = m$} {
            \Return $t, d + n * k_1, k + n * k_1$
        }
        \Return $t, \nan, \nan$
    }
    \BlankLine
    \ElseIf {$e = t_1 \in T$} {
        \If {$t \neq \nobasetag$ and $d \neq \nan$} {
            mark $t_1$ as fixed on $t$ with distance $d$ \;
            \Return $t, d, k$
        }
        \Return $t_1, 0, k$
    }
}

\vspace{1em}
\caption{Fixed tags optimization.}\label{alg_fixed_tags}
\end{algorithm}
\vspace{2em}

\columnbreak
\Indm\Indm

\includegraphics[width=\linewidth]{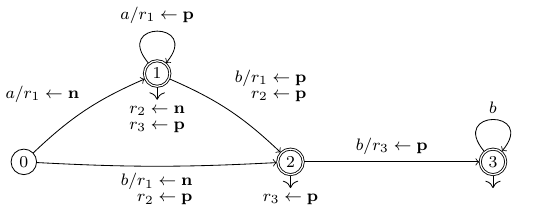}
\vspace{1em}
\captionof{figure}{
Optimized TDFA with fixed tags
$t_1 \leftarrow (\mathbf{n} \text{ if } t_2 = \mathbf{n} \text{ else } t_{2} - 1)$ and $t_3 \leftarrow (t_5 - 1)$. \\
Tags $t_2$, $t_4$, $t_5$ correspond to final registers $r_1$, $r_2$, $r_3$.
}\label{fig:tdfa_fixopt}

\end{multicols}
\medskip

Figure \ref{fig:tdfa_fixopt} shows the effect of fixed tags in addition to other optimizations on figure \ref{fig:tdfa_regopt}:
\medskip

\begin{itemize}

\item[\ding{212}]
In the example RE $(1a2)^*3(a|4b)5b^*$ tags $t_1$ and $t_2$ are within one symbol from each other,
so the value of $t_1$ can be computed as nil if $t_2$ is nil, or $t_2 - 1$ otherwise.
Likewise $t_3$ can be computed as $t_5 - 1$ (although there are multiple different paths through $(a|4b)$, they all have the same length).
\medskip

\item[\ding{212}]
Fixed tags are identified at the RE level and excluded from TNFA construction and determinization.
There are no registers and register operations associated with $t_1$ and $t_3$,
except for computing their values from the base tags $t_2$ and $t_5$ at the end of match.
\medskip

\end{itemize}

\FloatBarrier

\section{Multi-pass TDFA}\label{section_multipass}

TDFA with registers described in section \ref{section_tdfa}
are well suited for ahead-of-time determinization (e.g. in lexer generators)
when one can spend considerable time on optimizations in order to emit a more efficient TDFA.
However, in RE libraries the overhead on determinization and optimizations is included in the run time,
therefore it is desirable to reduce TDFA construction time
(although the overhead may be amortized if a RE is compiled once and used to match many strings).
\medskip

Another concern is the density of submatch information in a RE.
TDFA with registers are perfect for \emph{sparse} submatch extraction,
when the number of tags is small compared to the size of RE
and the runtime performance is expected be close to an ordinary DFA.
However, if a RE contains many tags (in the extreme, if every subexpression is tagged)
then transitions get cluttered with operations,
making TDFA execution very slow.
Moreover, the optimizations described in section \ref{section_impl} become problematic
due to the size of liveness and interference tables.
\medskip

Multi-pass TDFA address these issues:
they reduce TDFA construction time and they are better suited to dense submatch extraction.
The main difference with canonical TDFA is that multi-pass TDFA have no register operations.
Instead, as the name suggests, they have multiple passes:
a \emph{forward} pass that matches the input string and records a sequence of TDFA states,
and one or more \emph{backward} passes that iterate through the recorded states and collect submatch information
(one backward pass is sufficient, but an extra pass may be used e.g. to estimate and preallocate memory for the results).
The representation of submatch results may vary and affects only the backward pass(es);
forward pass is the same for all representations.
\medskip

\begin{algorithm}[b!] \DontPrintSemicolon \SetKwProg{Fn}{}{}{} \SetAlgoInsideSkip{medskip}
\begin{multicols}{2}
\setstretch{0.9}
\small

\Indm

\nonl\Fn {$\underline{unique \Xund origins \big( C \big)} \smallskip$} {
    $U:$ mapping from TNFA states in $C$ to integers \;
    $i = 0$ \;
    \BlankLine
    \For {each unique origin $o$ in $C$} {
        \For {each $(q, o', \Xund, \Xund, \Xund)$ in $C$ such that $o' = o$} {
            $U[q] = i$ \;
        }
        $i = i + 1$ \;
    }
    \BlankLine
    \Return $U$ \;
}

\vspace{2em}

\nonl\Fn {$\underline{match \big(
        \XF, \;
        a_1 \hdots a_n \big)} \smallskip$} {
    $V = \{ s_0 \}$ \;
    \For {$k = \overline{1, n}$} {
        \If {$s = \delta(s, a_k)$ is defined} {
            append $s$ to $V$ \;
        } \lElse {
            \Return $\varnothing$
        }
    }
    \Return $V$ \;
}

\vspace{2em}

\nonl\Fn {$\underline{extract \Xund o\!f\!\!f\!sets \big(
        \XF, \;
        a_1 \hdots a_n, \;
        s_0 \hdots s_n \big)} \smallskip$} {
    $E = \{ \varnothing \}_{i=1}^{|T|}$ (no value for each tag) \;
    $(i, h) = \varphi(s_n)$ \;
    $k = n$ \;
    \BlankLine
    \While {$true$} {
        \For {tag $t$ in $h$ in reverse order} {
            \If {$t > 0$ and $E[t] = \varnothing$} {
                $E[t] = k$
            } \ElseIf {$E[-t] = \varnothing$} {
                $E[-t] = -1$
                %\For {all tags $t'$ nested under $t$} {
                %    $E[t'] = -1$
                %}
            }
        }
        %\BlankLine
        \lIf {$k = 0$} { \bf{break} }
        %\BlankLine
        $(\Xund, B) = \delta(s_{k-1}, a_k)$ \;
        $(i, h) = B[i]$ \;
        $k = k - 1$ \;
    }
    \BlankLine
    \Return $E$ \;
}

\vfill\null

\columnbreak

\nonl\Fn {$\underline{construct \Xund backlinks \big( C, U, U' \big)} \smallskip$} {
    $B:$ backlink array of size $|range(U')|$ \;
    \BlankLine
    \For {each $(q, o, \Xund, h, \Xund)$ in $C$} {
        $i = U'[q]$ \;
        \If {$B[i]$ is undefined} {
            $B[i] = (U[o], h)$ \;
        }
    }
    \BlankLine
    \Return $B$ \;
}

\vspace{2em}

\nonl\Fn {$\underline{extract \Xund tstring \big(
        \XF, \;
        a_1 \hdots a_n, \;
        s_0 \hdots s_n \big)} \smallskip$} {
    $(i, h) = \varphi(s_n)$ \;
    $x = h$ \;
    \For {$k = \overline{n, 1}$} {
        $(\Xund, B) = \delta(s_{k-1}, a_k)$ \;
        $(i, h) = B[i]$ \;
        $x = h \cdot a_k \cdot x$ \;
    }
    \Return $x$ \;
}

\vspace{2em}

\nonl\Fn {$\underline{extract \Xund o\!f\!\!f\!set \Xund lists \big(
        \XF, \;
        a_1 \hdots a_n, \;
        s_0 \hdots s_n \big)} \smallskip$} {
    $E = \{ \{\} \}_{i=1}^{|T|}$ (empty list for each tag) \;
    $(i, h) = \varphi(s_n)$ \;
    $k = n$ \;
    \BlankLine
    \While {$true$} {
        \For {tag $t$ in $h$ in reverse order} {
            \If {$t > 0$} {
                prepend $k$ to $E[t]$
            } \Else {
                %\For {all tags $t'$ nested under $t$} {
                %    prepend $-1$ to $E[t']$
                %}
                prepend $-1$ to $E[-t]$
            }
        }
        %\BlankLine
        \lIf {$k = 0$} { \bf{break} }
        %\BlankLine
        $(\Xund, B) = \delta(s_{k-1}, a_k)$ \;
        $(i, h) = B[i]$ \;
        $k = k - 1$ \;
    }
    \BlankLine
    \Return $E$ \;
}

\vfill\null

\end{multicols}
\vspace{1em}
\caption{Backlink construction and matching with multi-pass TDFA $\XF = (\Sigma, T, S, S_f, s_0, \delta, \varphi)$.}\label{alg_multipass}
\end{algorithm}

Multi-pass TDFA construction differs from algorithm \ref{alg_tdfa} in section \ref{section_tdfa} in a few ways.
Recall that a closure $C$ corresponds to a TDFA transition between states $s$ and $s'$.
For multi-pass TDFA closure configurations are extended to five components $(q, o, r, h, l)$
where the new component $o$ is the \emph{origin} TNFA state in $s$ that leads to state $q$ in $s'$,
and the remaining components are as in algorithm \ref{alg_tdfa}.
Origins are needed to trace back the matching TNFA path from a sequence of TDFA states.
Path fragments corresponding to closure configurations are represented with \emph{backlinks},
and every TDFA transition is associated with a backlink array.
A backlink is a pair $(i, h)$ where $i$ is an index in backlink arrays on preceding transitions,
and $h$ is a tag sequence corresponding to the $h$-component of a configuration.
Backlinks on transitions do not map one-to-one to configurations,
because in TDFA(1) contrary to TDFA(0) configurations with identical origins have identical $h$-components (inherited from the lookahead tags),
resulting in identical backlinks.
To deduplicate such backlinks, $unique \Xund origins$ in algorithm \ref{alg_multipass}
creates a per-state mapping from origin state to a unique origin index.
Transition function is defined as $\delta(s, a) = (s', B)$ where $B = construct \Xund backlinks (C, U, U')$
and $U$, $U'$ are the unique origin mappings for $s$ and $s'$ respectively.
Final TDFA states are associated with a single backlink,
and the final function is defined as $\varphi(s) = (i, l)$
where $i$ is the unique origin index of the final state $q_f$ in $C$
and $l$ is the lookahead tag sequence ($l$-component of the final configuration).
The resulting TDFA has no registers or register operations;
the $R$ and $R_f$ components are removed from TDFA, and
functions $transition \Xund regops$, $final \Xund regops$ and their dependencies in algorithm \ref{alg_tdfa} are not needed.
%
%TDFA construction --- negative tags,
\medskip

Algorithm \ref{alg_multipass} shows matching with a multi-pass TDFA.
The forward pass is defined by the function $match$,
which executes TDFA on a string $a_1 \hdots a_n$ and returns the matching sequence of TDFA states $s_0 \hdots s_n$ (or $\varnothing$ on failure).
Backward pass depends on the representation of submatch results;
we provide three variants for offsets, offset lists and tagged strings.
In each case the backward pass follows a sequence of backlinks from the final state to the initial state.
Function $extract \Xund o\!f\!\!f\!sets$ extracts the last offset for each tag
and avoids overwriting it by initializing all offsets to $\varnothing$ and checking each offset before writing.
Function $extract \Xund o\!f\!\!f\!set \Xund lists$ is similar, but it collects offsets into lists.
Function $extract \Xund tstring$ concatenates the $h$-components of backlinks interleaved with input symbols
(this representation can be used to reconstruct a full parse tree, see \cite{BorTro19} section 6).
\medskip

In practice we found that the following details affect performance of algorithm \ref{alg_multipass}.
The $h$-components of backlinks should be stored as arrays which allow fast access to individual tags,
rather than linked lists packed in a prefix tree (the latter representation was used for multi-valued tags in section \ref{section_impl}).
The forward pass should record references to backlink arrays instead of TDFA states in order to reduce the number of indirections and lookups.
For tagged strings a separate backward pass may be used to estimate the amount of space for the resulting string and preallocate it.
If tags in a RE have \emph{nested} structure (e.g. in the case of POSIX capturing groups)
then negative transition should be added only for the topmost tag of a subexpression (as described in \cite{BorTro19} section 9)
rather than for all nested tags (as described in algorithm \ref{alg_tnfa}).
The mapping from a tag to its nested tags should be stored separately and used during matching (as in \cite{BorTro19} section 6).
\medskip

\begin{figure}[b]
\includegraphics[width=\linewidth]{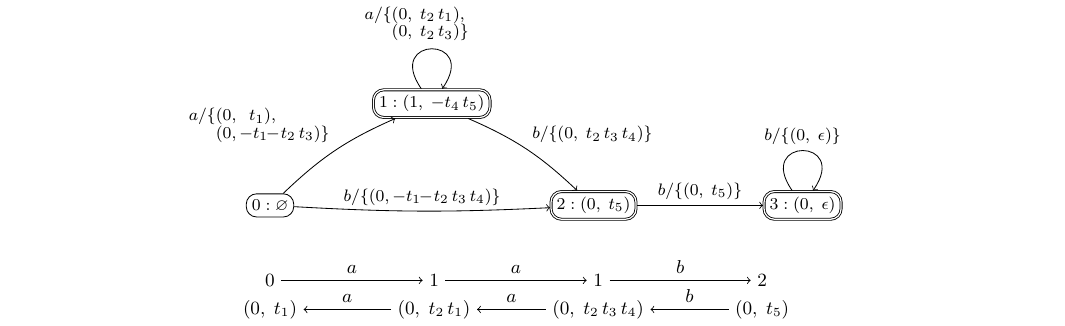}
\vspace{0.5em}
\captionof{figure}{
Multi-pass TDFA for RE $(1a2)^*3(a|4b)5b^*$ matching string $aab$.
}\label{fig:tdfa_multipass}
\end{figure}
\medskip

Figure \ref{fig:tdfa_multipass} shows multi-pass TDFA for the running example (compare it with figure \ref{fig:tdfa}):
\medskip

\begin{itemize}

\item[\ding{212}]
TDFA transition $0 \!\rightarrow\! 1$ has two backlinks
because the five configurations in state $1$ originate in two TNFA states $2$ and $9$.
Likewise transition $1 \!\rightarrow\! 1$ has two backlinks corresponding to origins $2$ and $9$,
transitions $0 \!\rightarrow\! 2$ and $1 \!\rightarrow\! 2$ have one backlink corresponding to origin $12$,
and transitions $2 \!\rightarrow\! 3$ and $3 \!\rightarrow\! 3$ have one backlink corresponding to origin $15$.
\medskip

\item[\ding{212}]
Final states $1$, $2$ and $3$ have a backlink corresponding to the final configuration with TNFA state $17$.
\medskip

\item[\ding{212}]
The $i$-component of each backlink equals to the unique origin index of the configuration $o$-component.
For example, both backlinks on transition $1 \!\rightarrow\! 1$ have $i=0$ because
their configurations (in the shadow TDFA state mapped to state $1$ on figure \ref{fig:tdfa}) have origins $2$ and $9$ in TDFA state $1$,
which have the same unique origin index $0$ (because they both have origin $2$ in TDFA state $0$).
Consequently, both backlinks on transition $1 \!\rightarrow\! 1$ connect to the first backlink on transitions $0 \!\rightarrow\! 1$ and $1 \!\rightarrow\! 1$.
On the other hand, the final backlink in state $1$ connects to the second one.
\medskip

\item[\ding{212}]
The sequence of TDFA states matching $aab$ is $0 \!\rightarrow\! 1 \!\rightarrow\! 1 \!\rightarrow\! 2$,
and backlinks can be traced back from the final backlink in state $2$
using $i$-component as index in backlink arrays.
\medskip

\item[\ding{212}]
Submatch results for string $aab$ are as follows.
Single offsets: $t_1\!=\!1$, $t_2\!=\!2$, $t_3\!=\!2$, $t_4\!=\!2$, $t_5\!=\!3$.
Offset lists: $t_1\!=\!\{0,1\}$, $t_2\!=\!\{1,2\}$, $t_3\!=\!\{2\}$, $t_4\!=\!\{2\}$, $t_5\!=\!\{3\}$.
Tagged string: $1\,a\,2\,1\,a\,2\,3\,4\,b\,5$.
\medskip

\end{itemize}

\FloatBarrier

\section{Evaluation}\label{section_evaluation}

\begin{figure}[t!]
\includegraphics[width=\linewidth]{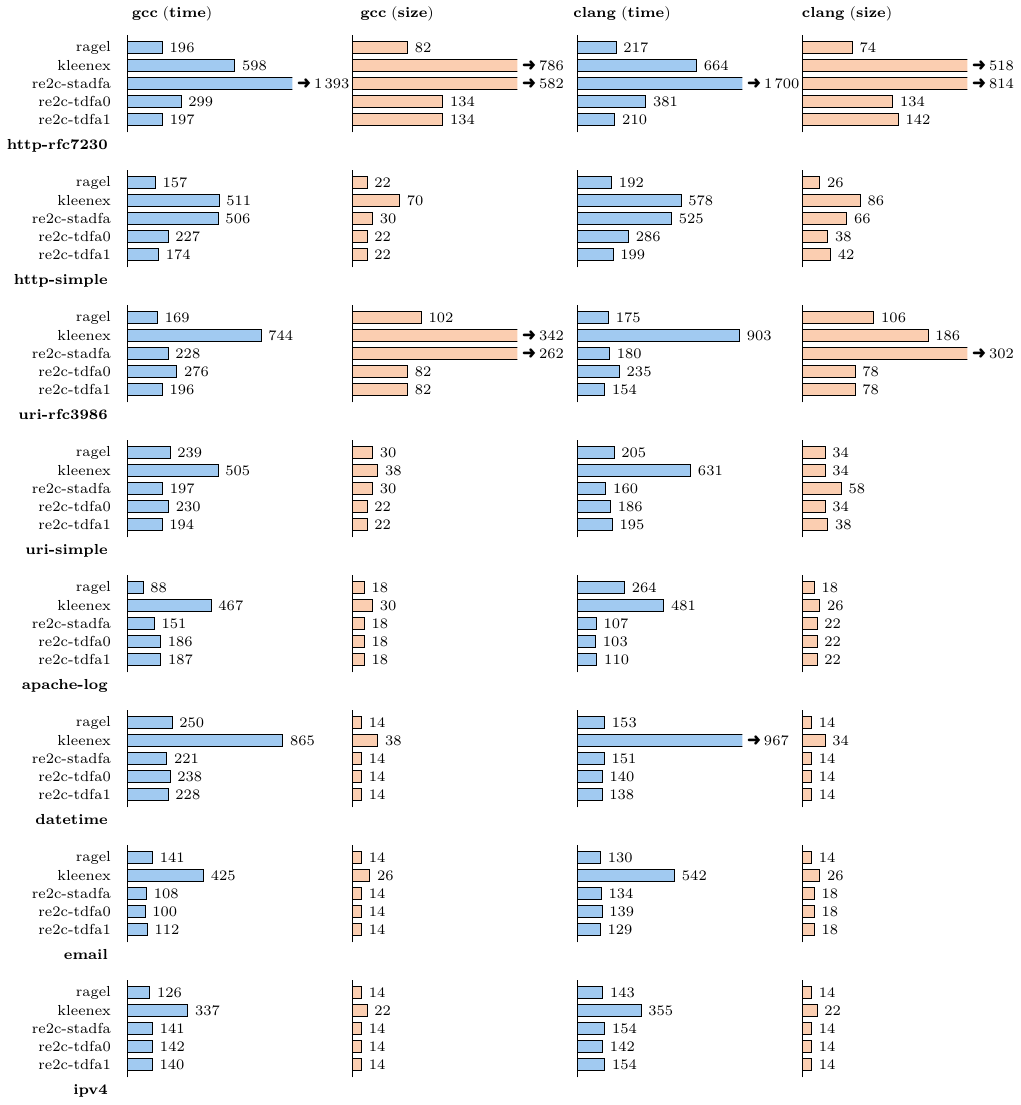}
\captionof{figure}{
Benchmarks for AOT determinization, real-world REs.
}\label{fig:benchmark_dfa_aot_main}
\medskip
\end{figure}

\begin{figure}[t!]
\includegraphics[width=\linewidth]{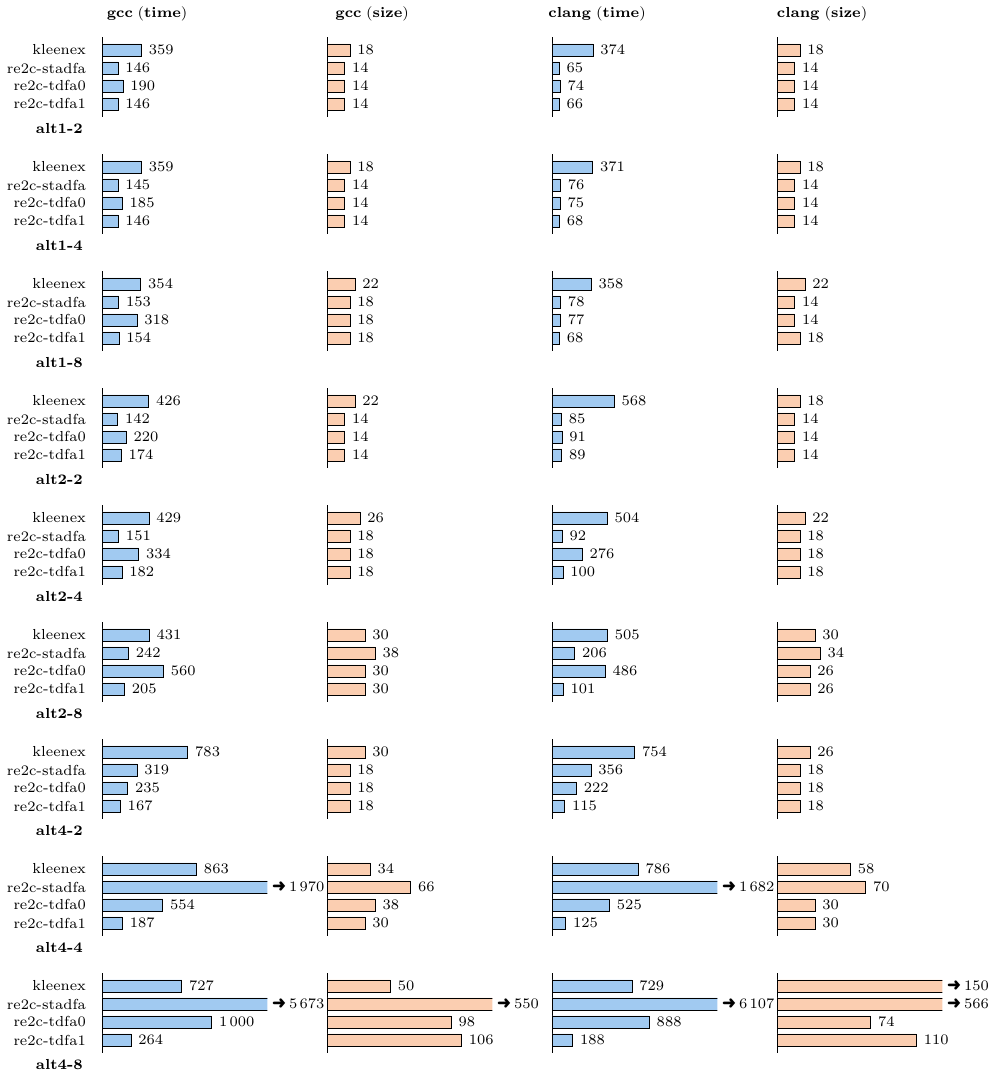}
\captionof{figure}{
Benchmarks for AOT determinization, artificial REs with alternative.
}\label{fig:benchmark_dfa_aot_alt}
\medskip
\end{figure}

\begin{figure}[t!]
\includegraphics[width=\linewidth]{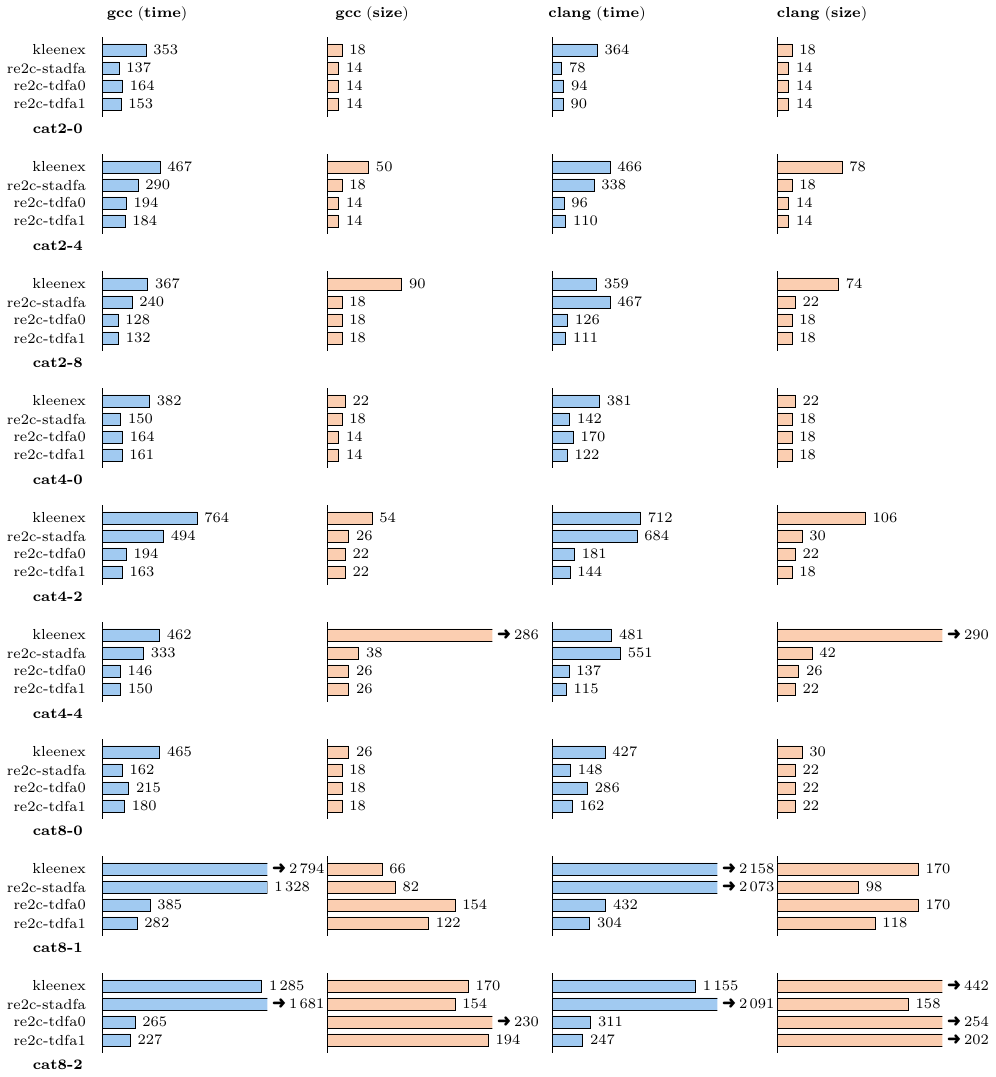}
\captionof{figure}{
Benchmarks for AOT determinization, artificial REs with concatenation.
}\label{fig:benchmark_dfa_aot_cat}
\end{figure}

\begin{figure}[t!]
\includegraphics[width=\linewidth]{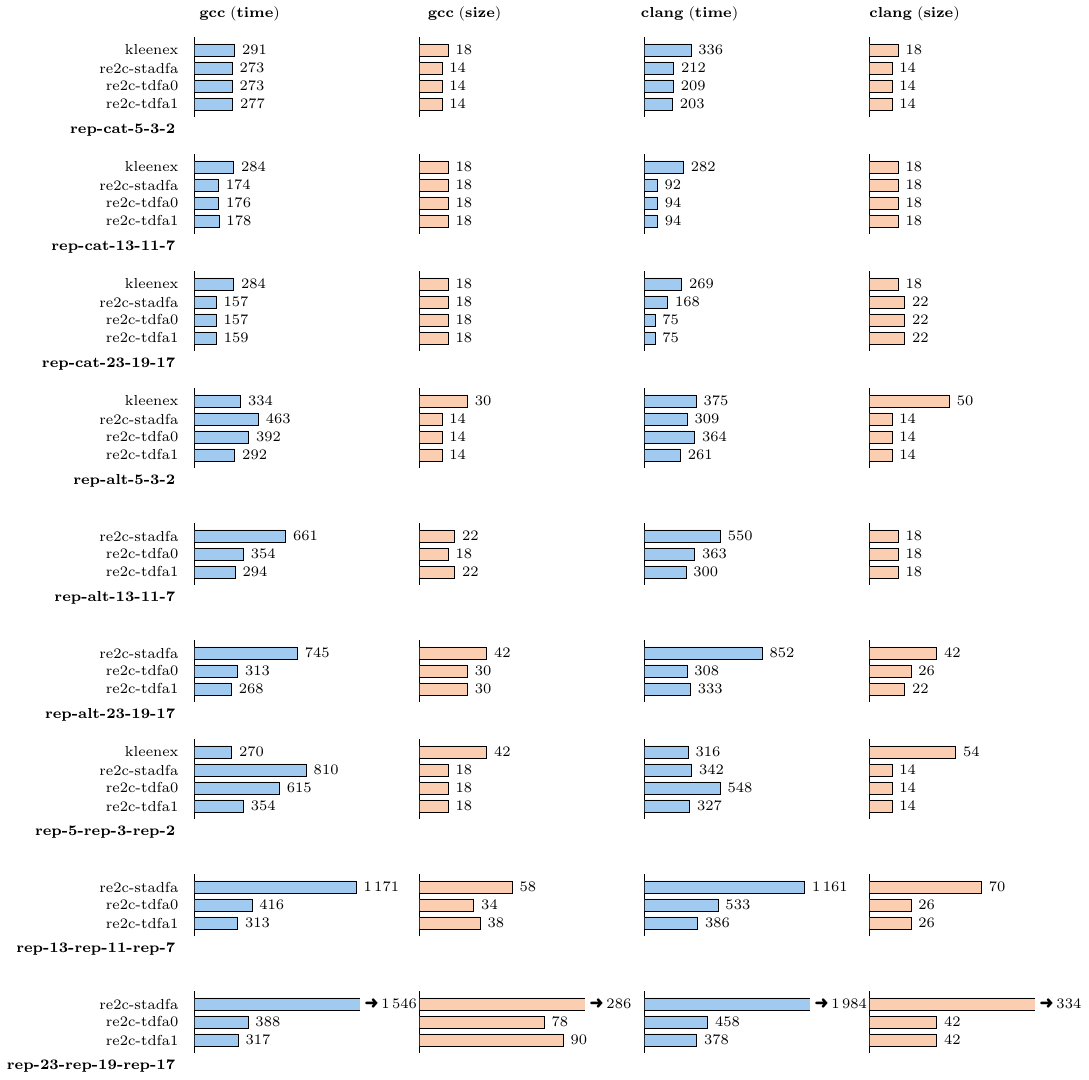}
\captionof{figure}{
Benchmarks for AOT determinization, artificial REs with repetition.
}\label{fig:benchmark_dfa_aot_rep}
\end{figure}

\begin{figure}[t!]
\includegraphics[width=\linewidth]{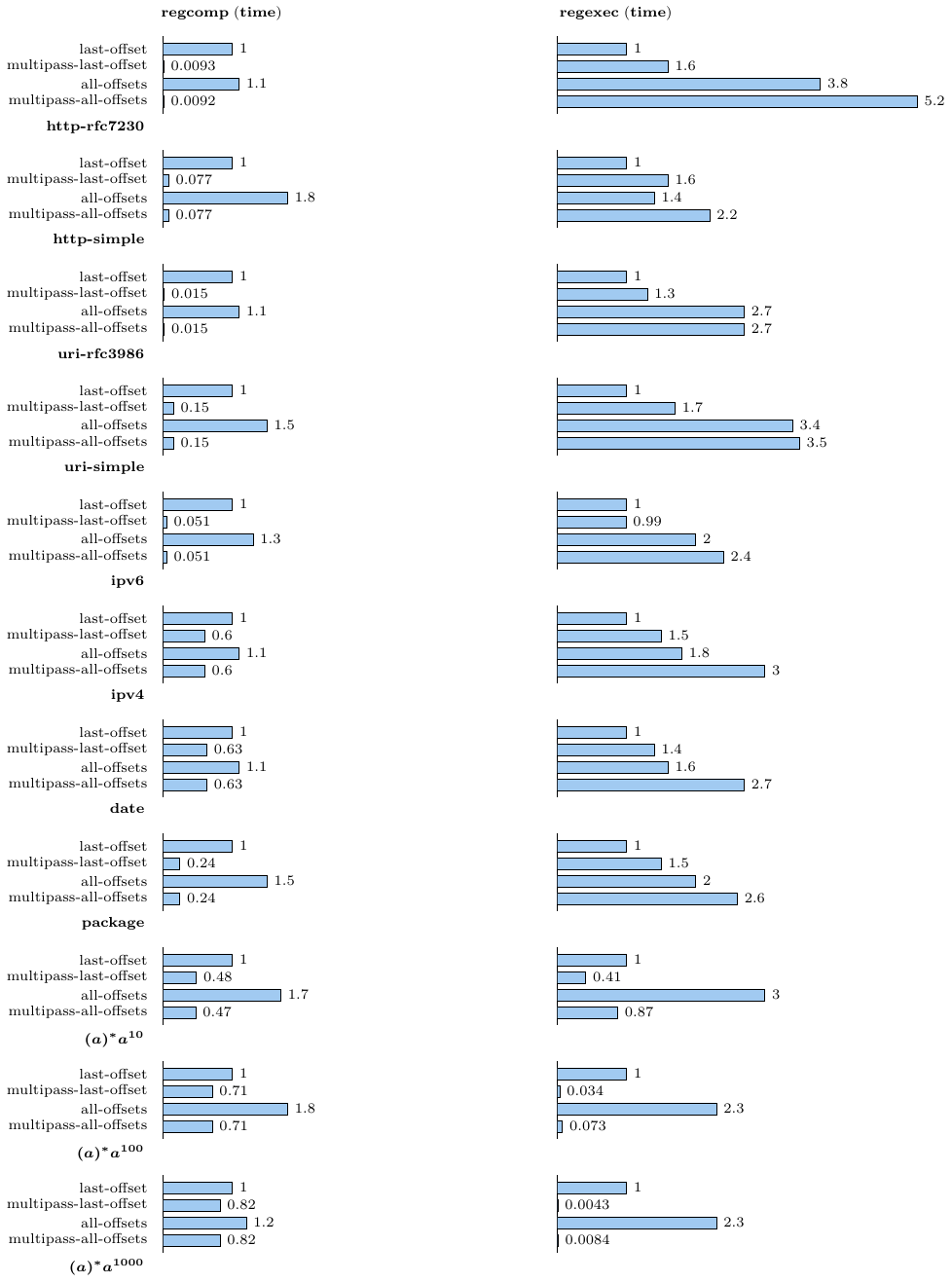}
\captionof{figure}{
Benchmarks for JIT determinization, C++ (\texttt{regcomp}/\texttt{regexec} time, relative).
}\label{fig:benchmark_dfa_jit}
\end{figure}

\begin{figure}[t!]
\includegraphics[width=\linewidth]{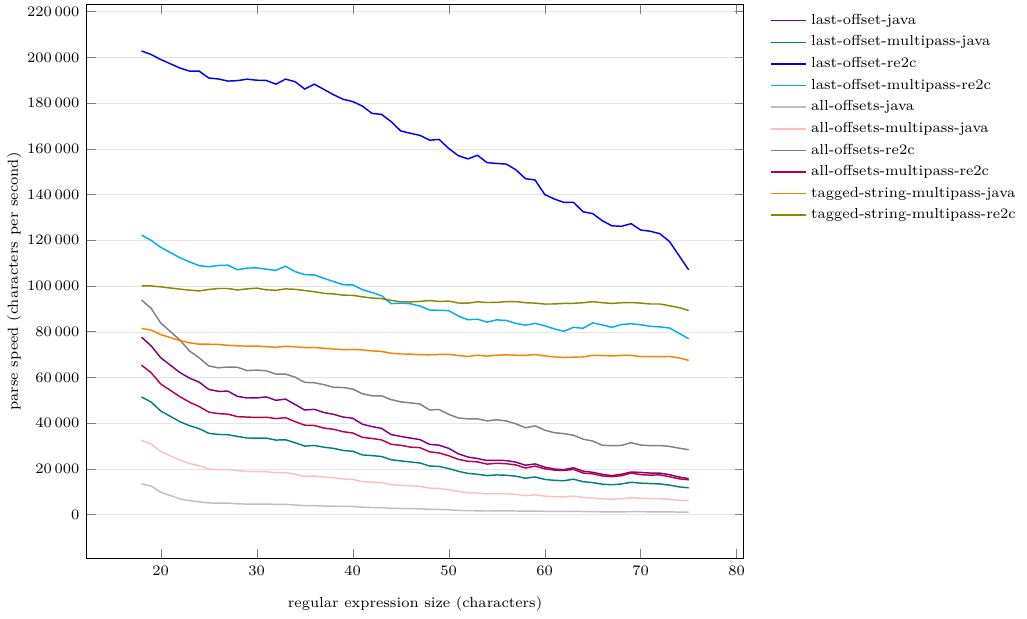}
\captionof{figure}{
Benchmarks for JIT determinization, Java, sparse tags (\texttt{regexec} speed).
}\label{fig:benchmark_java_sparse}
\end{figure}

\begin{figure}[b!]
\includegraphics[width=\linewidth]{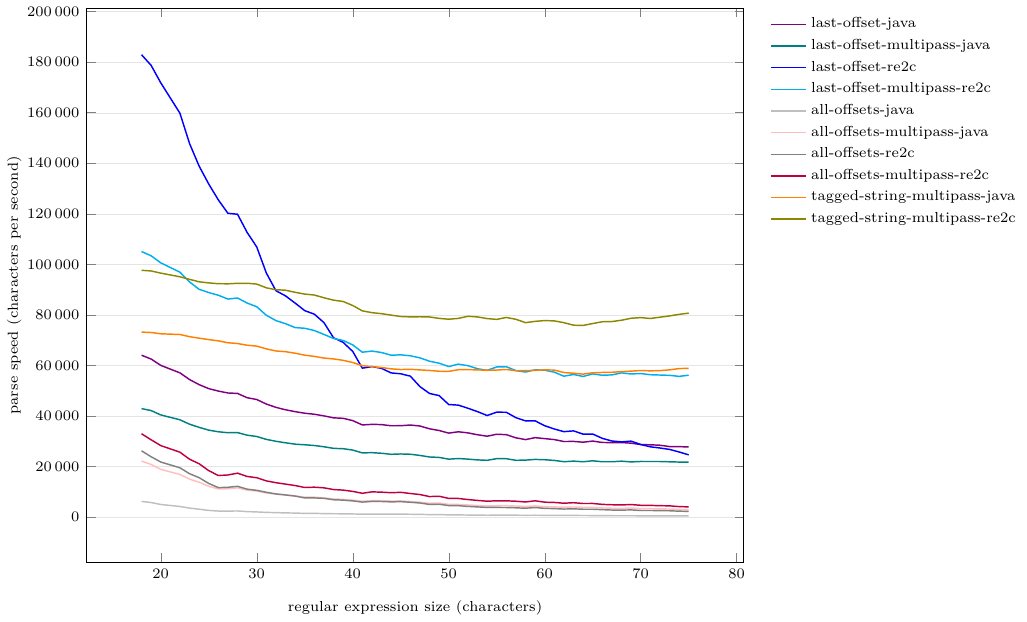}
\captionof{figure}{
Benchmarks for JIT determinization, Java, full parsing (\texttt{regexec} speed).
}\label{fig:benchmark_java_dense}
\end{figure}

In this section we evaluate TDFA performance in practice and compare it to other algorithms.
We present three groups of benchmarks that cover different settings and show different aspects of the algorithm:
\medskip

\begin{enumerate}%[leftmargin=1.2em]

\item
\textbf{AOT determinization}
(figures \ref{fig:benchmark_dfa_aot_main}, \ref{fig:benchmark_dfa_aot_alt}, \ref{fig:benchmark_dfa_aot_cat}, \ref{fig:benchmark_dfa_aot_rep}).
We compare three lexer generators: RE2C \cite{RE2C}, Ragel \cite{Ragel} and Kleenex \cite{Kleenex}
that are based on different types of deterministic automata.
All of them generate optimized C code, which is further compiled to binary by GCC and Clang.
The generated programs do string rewriting: they read 100MB of input text and insert markers at submatch extraction points.
These benchmarks use leftmost-greedy disambiguation.
The following automata are compared:
\medskip

\begin{itemize}

\item[$\bullet$]
\textbf{TDFA(1)}, the algorithm described by Trafimovich in \cite{Tro17} and presented in this paper.
It is implemented in RE2C with the optimizations described in section \ref{section_impl}.
\medskip

\item[$\bullet$]
\textbf{TDFA(0)}, the original algorithm described by Laurikari in \cite{Lau00}.
Contrary to TDFA(1) that use one-symbol lookahead, TDFA(0) do not use lookahead:
the two types of automata are called so by analogy with LR(1) and LR(0).
TDFA(0) apply register operations to the incoming transition,
while TDFA(1) split them on the lookahead symbol and apply them to the outgoing transitions, which reduces the number of tag conflicts.
As a consequence, TDFA(0) typically require more registers and copy operations, which makes them slower than TDFA(1);
see \cite{Tro17} for a detailed comparison.
TDFA(0) algorithm is also implemented in RE2C and benefits from the same optimizations as TDFA(1).
\medskip

\item[$\bullet$]
\textbf{StaDFA}, the algorithm described by Chowdhury in \cite{Cho18},
with a few modifications of our own that were necessary for correctness.
It is very similar to TDFA, but the automata have register operations in states rather than on transitions
(which implies that staDFA do not use lookahead).
The algorithm is implemented in RE2C and uses the same optimizations as TDFA(1) and TDFA(0).
\medskip

\item[$\bullet$]
\textbf{DSSTs}, the algorithm described by Grathwohl in \cite{Gra15}.
DSSTs stands for Deterministic Streaming String Transducers; these are more distant relatives to TDFA,
better suited to string rewriting and full parsing.
DSST states contain path trees constructed by the $\epsilon$-closure,
while TDFA states contain similar information decomposed into register tables and lookahead tags.
DSST registers contain fragments of strings over the output alphabet (the analogue of our tagged strings).
Register operations on transitions concatenate and move string fragments.
DSSTs are implemented in Kleenex.
\medskip

\item[$\bullet$]
Ordinary \textbf{DFA} with ad-hoc user-defined actions
and manual conflict resolution via precedence operators,
implemented in Ragel.
This approach is fast, but it has correctness issues:
in some cases it is impossible to resolve the conflicts between actions by preferring one action over the other;
instead, it is necessary to keep both actions until more input is consumed and non-determinism is resolved.
But this is also impossible, as the actions modify the same shared state (e.g. set the same local variables).
An action may conflict with itself on different transitions due to non-determinism.
\medskip

\end{itemize}

\item
\textbf{JIT determinization, C++}
(figure \ref{fig:benchmark_dfa_jit}).
These benchmarks compare \textbf{TDFA(1)} and \textbf{multi-pass TDFA(1)} presented in section \ref{section_multipass}
in the case of \textbf{single offsets} and \textbf{offset lists}.
Both algorithms are implemented in a C++ library based on RE2C.
These benchmarks use POSIX disambiguation.
\medskip

\item
\textbf{JIT determinization, Java}
(figures \ref{fig:benchmark_java_sparse}, \ref{fig:benchmark_java_dense}).
We compare two independent implementations:
one in pure Java (by Borsotti),
and one in C++ via JNI, based on RE2C (by Trafimovich),
both published as part of the RE2C repository \cite{RE2CJava}.
These benchmarks compare \textbf{TDFA(1)} and \textbf{multi-pass TDFA(1)}
in the case of \textbf{single offsets}, \textbf{offset lists} and \textbf{tagged strings},
and with different submatch density: \textbf{sparse tags} and \textbf{full parsing}
(where every subexpression in a RE is tagged).
They use POSIX disambiguation.
\medskip

\end{enumerate}

Hardware specifications:
Intel Core i7-8750H CPU with
32 KiB L1 data cache,
32 KiB L1 instruction cache,
256 KiB L2 cache,
9216 KiB L3 cache,
32 GiB RAM.
Software versions:
RE2C 3.0,
Ragel 7.0.4,
Kleenex built from Git at commit d474c60,
GCC 11.2.0,
Clang 13.0.0,
OpenJDK 17.0.1.
\medskip

\pagebreak

Figure \ref{fig:benchmark_dfa_aot_main} shows benchmark results for AOT determinization
in the case of real-world REs that are likely to be used in practice:
HTTP message headers, URI, Apache logs, date, email addresses and IP addresses.
REs vary from very large and complex to small and simple;
the number of tags in REs varies accordingly.
\medskip

The main conclusions are:
\medskip

\begin{itemize}

\item[$\bullet$]
TDFA(1) and ordinary DFA are close in size and speed (the result often depends on GCC/Clang).
\medskip

\item[$\bullet$]
In simple cases staDFA and TDFA(0) are on par with TDFA(1),
but in complex cases they are considerably slower and larger,
and staDFA can get extremely large.
\medskip

\item[$\bullet$]
DSSTs are generally slower and larger in most of the cases.

\end{itemize}

\pagebreak

Figure \ref{fig:benchmark_dfa_aot_alt} shows benchmark results for AOT determinization
in the case of artificial REs with emphasis on alternative, in series of increasing size, complexity and the number of tags.
Ordinary DFA are excluded because Ragel's ad-hoc disambiguation operators do not allow to implement all cases correctly.
Conclusions:
\medskip

\begin{itemize}

\item[$\bullet$]
TDFA(1) perform better than other algorithms.
\medskip

\item[$\bullet$]
TDFA(0) are generally slower than TDFA(1); the difference grows with RE size.
\medskip

\item[$\bullet$]
StaDFA are close to TDFA(1) on small REs, but they degrade on large REs in both size and speed.
\medskip

\item[$\bullet$]
DSSTs are generally slower and almost always larger than TDFA(1).

\end{itemize}

\pagebreak

Figure \ref{fig:benchmark_dfa_aot_cat} shows benchmark results for AOT determinization
in the case of artificial REs with emphasis on concatenation, in series of increasing size, complexity and the number of tags.
Ordinary DFA are excluded because Ragel's ad-hoc disambiguation operators do not allow to implement all cases correctly.
Conclusions:
\medskip

\begin{itemize}

\item[$\bullet$]
TDFA(1) perform better than other algorithms.
\medskip

\item[$\bullet$]
TDFA(0) are slower than TDFA(1), but the difference is not radical.
\medskip

\item[$\bullet$]
StaDFA are slower than TDFA(1) on small REs, and the difference gets radical with RE size.
\medskip

\item[$\bullet$]
DSSTs are generally slower and almost always larger than TDFA(1).

\end{itemize}

\pagebreak

Figure \ref{fig:benchmark_dfa_aot_rep} shows benchmark results for AOT determinization
in the case of artificial REs with emphasis on repetition, in series of increasing size, complexity and the number of tags.
Ordinary DFA are excluded because Ragel's ad-hoc disambiguation operators do not allow to implement all cases correctly,
and DSSTs are excluded in cases where they get too large to be compiled.
Conclusions:
\medskip

\begin{itemize}

\item[$\bullet$]
TDFA(1) perform better than other algorithms.
\medskip

\item[$\bullet$]
TDFA(0) are slower than TDFA(1), but the difference is not radical.
\medskip

\item[$\bullet$]
StaDFA are slower and larger than TDFA(1) on small REs, and the difference gets radical with RE size.
\medskip

\item[$\bullet$]
DSSTs are generally larger than TDFA(1), and the difference gets extreme with RE size.
\medskip

\end{itemize}

\pagebreak

\FloatBarrier

\FloatBarrier

Figure \ref{fig:benchmark_dfa_jit} shows benchmark results for JIT determinization, C++.
Time is shown relative to the first row.
There are two groups of benchmarks: real-world REs and artificial REs $(a)^* a^{10^k}$ for $k \in \{1, 2, 3 \}$.
Conclusions:
\medskip

\begin{itemize}

\item[$\bullet$]
Compilation is predictably slower for TDFA than for multi-pass TDFA for both groups,
as multi-pass TDFA do not need register actions and subsequent register optimizations.
\medskip

\item[$\bullet$]
Execution time differs for the two groups:
for real-world benchmarks TDFA are generally faster than multi-pass TDFA,
while for artificial benchmarks TDFA are much slower than multi-pass TDFA, and the difference grows with the size of RE.
In fact artificial REs demonstrate a pathological case for TDFA with register actions:
increasing $k$ results in increased degree of nondeterminism,
which requires more registers and copy operations in order to track all nondeterministic values.
High degree of nondeterminism is specific to some REs with counted repetition, as demonstrated in \cite{Tro17} (page 21).
\medskip

\item[$\bullet$]
The results are similar for single-offset and offset-list cases,
although the latter is predictably slower.
\medskip

\end{itemize}

\medskip

Figures \ref{fig:benchmark_java_sparse} and \ref{fig:benchmark_java_dense} show benchmark results for JIT determinization, Java,
in the case of sparse tags and full parsing respectively.
The plots show the dependence of matching speed on RE size.
Conclusions:
\medskip
\begin{itemize}

\item[$\bullet$]
Remarkably, the case of tagged strings with multi-pass TDFA is the only one that shows almost no degradation with RE size (the lines are almost horizontal).
This holds for both implementations.
\medskip

\item[$\bullet$]
TDFA with register actions (the RE2C implementation) is clearly the fastest algorithm in the case of sparse tags.
However, in the case of full parsing it either degrades faster than multi-pass TDFA (in the last-offset case),
or it is generally slower (in the offset-list case).
For pure-Java implementation multi-pass TDFA is almost always faster than TDFA with register actions.
\medskip

\end{itemize}

\section{Conclusions}\label{section_conclusions}

TDFA(1) are generally faster and smaller than other automata capable of submatch extraction.
%(TDFA(0), staDFA, DSST).
\medskip

Optimizations play a very important part in any performance-sensitive TDFA implementation
(compare the unoptimized TDFA on figure \ref{fig:tdfa} with the final optimized TDFA on figure \ref{fig:tdfa_fixopt}).
\medskip

The overhead on submatch extraction depends on tag density and degree of nondeterminism in a RE.
In the case of sparse tags with low nondeterminism TDFA with register actions are by far the fastest and have negligible difference compared to ordinary DFA.
In the case of high tag density (in the extreme, full parsing) or in the case of highly nondeterministic REs
multi-pass TDFA are more efficient.
\medskip

The overhead on submatch extraction depends on the representation of submatch results.
Tagged string extraction with multi-pass TDFA is the only algorithm that shows almost no degradation with RE size.
Extracting only the last offset is predictably faster than extracting all offsets
(fortunately, the choice is individual for each tag, so all offsets can be extracted only for a selected subset of tags).
\medskip

Multi-pass TDFA are better suited to JIT determinization than TDFA with register actions.
\medskip

\section{Future work}\label{section_future_work}

One very useful direction of future work is to find \emph{deterministic points} in a RE.
Often shifting a tag by a fixed number of characters in a concatenation subexpression can reduce its degree of nondeterminism
(the maximum number of registers in a single TDFA state needed to track all parallel versions of the same tag).
As a consequence, this means fewer registers and register operations.
For example, tag $t_1$ in $a^* 1 a^k a^*$ has nondeterminism degree $k$ and requires $2*k$ register operations,
while tag $t_2$ in $a^* a^k 2 a^*$ has degree is $1$ and only $1$ operation.
But tags $t_1$ and $t_2$ are within fixed distance of $k$ characters, so $t_1$ can be the computed as $t_2 - k$.
In other words, $t_2$ is a deterministic point for $t_1$.
Identifying such points in a RE would be a useful optimization.

\section*{Acknowledgments}

I want to thank my parents Vladimir and Elina,
my dearest friend and open source programmer Sergei,
my teachers Tatyana Leonidovna and Demian Vladimirovich
and the whole open source community.
And, of course, my coauthor Angelo who was the greatest inspiration and help in this work!
\null\hfill\textit{Ulya Trafimovich}

\pagebreak

\end{document}